\documentclass[pdflatex,sn-mathphys-num,iicol]{sn-jnl}% Math and Physical Sciences Numbered Reference Style
\usepackage{graphicx}%
\usepackage{multirow}%
\usepackage{adjustbox}
\def\blue#1{\textcolor{blue}{#1}}
\usepackage{amsmath,amssymb,amsfonts}%
\usepackage{bm}      % 支持向量粗体（如 \bm{x}）
\usepackage{amsthm}%
\usepackage[title]{appendix}%
\usepackage{xcolor}%
\usepackage{textcomp}%
\usepackage{manyfoot}%
\usepackage{booktabs}%
\usepackage{hyperref}
\usepackage{algorithm}%
\usepackage{algorithmic}
\usepackage{caption}
\usepackage{listings}%
\usepackage{xspace}%
\usepackage{array}    % 支持表格列宽调整
\usepackage{booktabs} % 支持 \toprule、\midrule、\bottomrule
\usepackage{multicol}
\usepackage{multirow}
\usepackage{makecell}

\makeatletter
\def\onedot{\ifx\@let@token.\else.\null\fi\xspace}
\makeatother
\def\eg{\emph{e.g}\onedot} 
\def\ie{\emph{i.e}\onedot} 
\newcommand{\sign}{\text{sign}}
\newcommand{\red}[1]{\textcolor{red}{#1}}
\usepackage{xcolor}

\definecolor{darkred}{RGB}{191, 0, 255}
\renewcommand{\blue}[1]{#1}
\colorlet{tablecolor}{black}
\theoremstyle{thmstyleone}%
\newtheorem{theorem}{Theorem}%  meant for continuous numbers
\newtheorem{proposition}{Proposition}
\theoremstyle{thmstyletwo}%

\theoremstyle{thmstylethree}%
\newtheorem{definition}{Definition}%

\begin{document}

\title[Article Title]{SWAP: Towards Copyright Auditing of Soft Prompts via Sequential Watermarking}


\author[1]{\fnm{Wenyuan} \sur{Yang}}\email{yangwy56@mail.sysu.edu.cn}
\equalcont{These authors contributed equally to this work.}

\author[2]{\fnm{Yichen} \sur{Sun}}\email{yichensun@zju.edu.cn}
\equalcont{These authors contributed equally to this work.}

\author[1]{\fnm{Changzheng} \sur{Chen}}\email{chenchzh23@mail2.sysu.edu.cn}

\author[2]{\fnm{Zhixuan} \sur{Chu}}\email{chuzhixuan@zju.edu.cn}

%\author[5]{\fnm{Zhifeng} \sur{Li}}\email{zhifeng0.li@gmail.com}

\author[3]{\fnm{Jiaheng} \sur{Zhang}}\email{jhzhang@nus.edu.sg}

\author*[4]{\fnm{Yiming} \sur{Li}}\email{liyiming.tech@gmail.com}

\author[4]{\fnm{Dacheng} \sur{Tao}}\email{dacheng.tao@gmail.com}

% \affil*[1]{\orgdiv{Department}, \orgname{Organization}, \orgaddress{\street{Street}, \city{City}, \postcode{100190}, \state{State}, \country{Country}}}
\affil[1]{\orgname{Sun Yat-sen University}, \orgaddress{ \city{Guangzhou}, \postcode{518107}, \country{China}}}

% \affil[3]{ \orgname{Key Laboratory of Cyberspace Security, Ministry of Education}, \orgaddress{ \city{Zhengzhou}, \postcode{450002}, \country{China}}}

% \affil[4]{ \orgname{Henan Key Laboratory of Cyberspace Situation Awareness}, \orgaddress{ \city{Zhengzhou}, \postcode{450002}, \country{China}}}

\affil[2]{\orgname{Zhejiang University}, \orgaddress{ \city{Hangzhou}, \postcode{310027}, \country{China}}}

\affil[3]{\orgname{National University of Singapore}, \orgaddress{ \postcode{119077}, \country{Singapore}}}

\affil[4]{\orgname{Nanyang Technological University}, \orgaddress{ \postcode{639798}, \country{Singapore}}}

\abstract{Large-scale vision-language models, especially CLIP, have demonstrated remarkable performance across diverse downstream tasks. Soft prompts, as carefully crafted modules that efficiently adapt vision–language models to specific tasks, necessitate effective copyright protection. In this paper, we investigate model copyright protection by auditing whether suspicious third-party models incorporate protected soft prompts. While this can be viewed as a special case of model ownership auditing, our analysis shows that existing techniques are ineffective due to prompt learning's unique characteristics. Non-intrusive auditing is inherently prone to false positives when independent models share similar data distributions with victim models. Intrusive approaches also fail: backdoor methods designed for CLIP cannot embed functional triggers, while extending traditional DNN backdoor techniques to prompt learning suffers from harmfulness and ambiguity challenges. We find that these failures in intrusive auditing stem from the same fundamental reason: watermarking operates within the same decision space as the primary task yet pursues opposing objectives. Motivated by these findings, we propose sequential watermarking for soft prompts (SWAP), which implants watermarks into a different and more complex space. SWAP encodes watermarks through a specific order of defender-specified out-of-distribution classes, inspired by the zero-shot prediction capability of CLIP. This watermark, which is embedded in a more complex space, keeps the original prediction label unchanged, making it less opposed to the primary task. \blue{We further design a hypothesis-test-guided verification protocol for SWAP and provide a theoretical analysis of when verification works. Extensive experiments on 11 datasets demonstrate SWAP's effectiveness, harmlessness, and robustness against potential attacks.}
}
\keywords{Model watermarking, Copyright protection, Prompt tuning, Vision-language model}



\maketitle
% \clearpage
\section{Introduction}\label{sec:intro}
%全文改成model ownership auditing

Vision-language models (VLMs), especially contrastive language-image pretraining (CLIP) \citep{radford2021learning}, have demonstrated remarkable capabilities across various tasks, including semantic segmentation, cross-modal retrieval, vision question answering, action recognition, and unsupervised semantic segmentation\citep{zhu2025weakclip,ventura2025learning,wang2025reclip++,wang2024clip}, with their ability to align visual and textual representations for effective zero-shot transfer learning. In particular, researchers have increasingly adopted soft prompting techniques \citep{zhou2022conditional,khattak2023maple,khattak2023self,bulat2024language} to enhance CLIP's performance through learned continuous vector representations that guide the model toward specific tasks without parameter modification. These learned soft prompts can be used as plug-and-play components, enabling their rapid adaptation to new domains and improving their performance on specialized tasks \citep{gao2024clip,zeng2025supplementary,jing2025animal}. In general, the training of these soft prompts requires substantial computational resources and domain-specific data (\eg, medical records or artworks), making them valuable intellectual property assets. In many cases, these soft prompts are public \citep{zhou2022conditional,zhou2022learning,khattak2023maple}. For example, developers may upload them to open-sourced platforms (\eg, GitHub and Hugging Face) and allow their usage for educational or research purposes; developers may also sell them and require that they cannot be sold again without permission. However, intellectual property concerns also arise due to their public nature, where adversaries may exploit these prompts for impermissible commercial purposes \citep{li2025rethinking}.

%Currently, model ownership verification (MOV)  is the most classical and effective method to protect the copyright of these public models. In general, MOV methods are categorized into two main classes: non-intrusive auditing methods and intrusive auditing methods. Non-intrusive auditing methods \citep{maini2021dataset,peng2022fingerprinting,dziedzic2022dataset,shokri2017membership,adv-tra,yang2022metafinger} first extract model's inherent yet distinctive features from the model by the use of methods such as class distances, adversarial perturbations, or decision trajectories. These extracted features are then used as a unique model fingerprint that can identify whether a suspicious model shares the same underlying learned patterns as the original model. Intrusive auditing method \citep{adi2018turning,gan2023towards,li2025move} actively embeds a unique, verifiable signature directly into a model's behavior. 

Currently, model ownership auditing stands as the most established and effective approach for safeguarding the copyright of publicly released models. Existing model ownership auditing methods can be categorized into two classes: non-intrusive auditing and intrusive auditing. Non-intrusive auditing methods \citep{maini2021dataset,peng2022fingerprinting,shao2025sok} extract inherent yet distinctive characteristics of a model, such as class-wise distances, adversarial perturbation responses, or decision trajectories, and use them to construct a unique fingerprint. This fingerprint serves as an identifier to determine whether a suspicious model preserves the same underlying learned patterns as the original one. Intrusive auditing methods \citep{wang2025sleepermark,ya2023towards,gan2023towards}, in contrast, rely on the intrusive injection of a developer-specific identifier into the model, thereby embedding a unique and verifiable signature into its behavior to enable direct ownership verification. A widely adopted intrusive approach is backdoor-based watermarking \citep{nguyen2021wanet,liu2025pre,liang2025vl}, which introduces specific external or artificial features during training. This process typically consists of two stages: model watermarking and ownership verification. In the watermarking stage, developers embed distinctive patterns into the model such that any unauthorized derivative inevitably inherits them. In the subsequent ownership verification stage, defenders determine whether a suspicious third-party model infringes on the original by extracting its potential watermark and comparing it against the owner-specified reference. Notably, ownership verification is often performed in a black-box setting \citep{guo2024zero,li2025reliable,shao2025explanation}, where the defender can only interact with the model through queries and observe its outputs, without access to source files or intermediate states (\eg, gradients). This constraint arises because third-party commercial models are commonly deployed as services accessible solely via APIs.

% Currently, model ownership verification (MOV) \cite{adi2018turning,gan2023towards,li2025move} is the most classical and effective method to protect the copyright of these public models. In general, MOV consists of two main stages: model watermarking and ownership verification. During model watermarking, model developers embed distinctive patterns that will be inherited by any unauthorized derivatives into their models. The subsequent ownership verification justifies whether (parts of) a suspicious third-party model are infringements by extracting its potential watermark and comparing it to an owner-specified one. In particular, the ownership verification is usually conducted in a \emph{black-box} manner \cite{guo2024zero,li2025reliable,shao2025explanation}, where defenders can only query the suspicious model and obtain its generation without accessing its source files and intermediate results (\eg, gradients). This is mainly because third-party commercial models usually provide services via their APIs. Consequently, backdoor-based watermarking \citep{zhang2018protecting,wang2024diagnosis, gan2023towards} has emerged as the predominant technique used in the first stage since they can implant distinctive (\eg, misclassification) behaviors as watermark patterns that will only appear on particular verification samples for stealthiness. 

Arguably, the copyright protection of soft prompts can be regarded as a special case of model ownership auditing, since their contribution is primarily manifested in the inference behavior of the specialized models they augment. \blue{However, unlike conventional model ownership auditing methods, which typically treats the entire model as the verification target, prompt ownership verification introduces new challenges due to the distinctive nature of prompt learning.} The most critical difficulty lies in the extremely limited parameter space of soft prompts (typically less than 0.1\% of the overall CLIP model), as the backbone architecture of CLIP is usually left intact to preserve its generalization ability in prompt learning. In this paper, we show that existing model ownership auditing techniques fail to address these unique challenges. We first reveal that non-intrusive auditing methods (\ie, fingerprinting) are susceptible to false positives, as models trained on datasets with similar distributions tend to converge toward analogous intrinsic features, consistent with findings in traditional image classification \cite{li2025move}. This problem is exacerbated in CLIP soft prompts, where the small-scale modifications leave the model's large-scale features essentially unchanged. Given these fundamental limitations, this paper turns to intrusive auditing, particularly backdoor-based watermarking, as a more viable path.

\begin{figure}
  \centering
\includegraphics[width=1\linewidth]{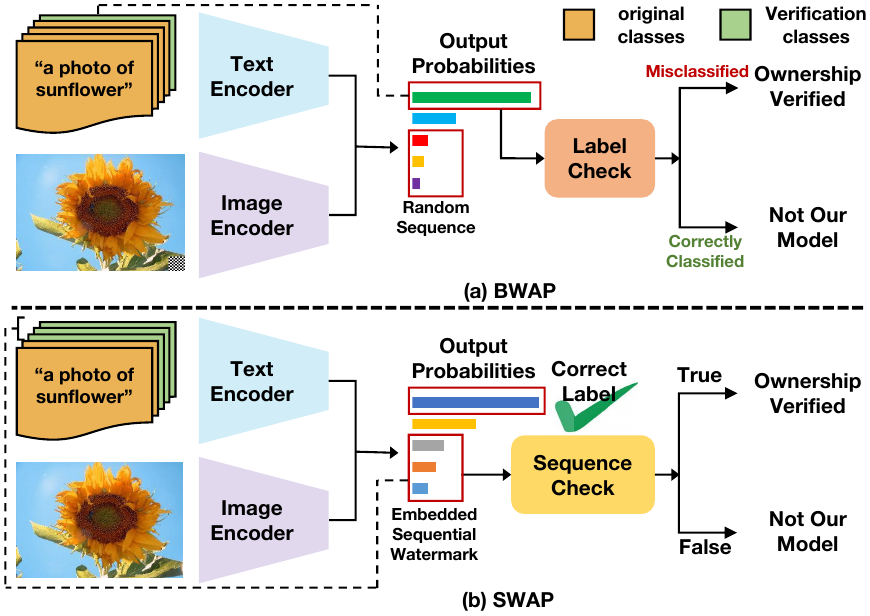}
%\vspace{-0.5em}
  \caption{The comparison between the backdoor-based watermarking scheme (\ie, BWAP) and our proposed SWAP for CLIP soft prompts. BWAP determines ownership through induced misclassification, which inevitably alters the model’s predictions. In contrast, SWAP verifies ownership by examining the sequential ordering of additional defender-specified classes rather than changing predictions. This design preserves the model’s utility while enabling reliable ownership verification.}
  \vspace{-1em}
  \label{comparison}
\end{figure}

Unfortunately, developing an effective backdoor-based auditing scheme for protecting soft prompts in CLIP models is non-trivial. Directly applying existing backdoor attacks against CLIP models, or adapting existing backdoor watermarking methods designed for traditional DNNs, both encounter inherent limitations. Specifically, the direct application of existing backdoor attacks against CLIP models to soft prompts fails to embed the watermark. We argue that this is mainly because these methods are designed to modify a massive number of model parameters to succeed. In contrast, soft prompts represent a parameter-efficient module comprising only a tiny fraction of the total CLIP parameters. \blue{The small parameter space is insufficient to make these poisoning-based methods (with limited poisoned samples) effective.} \blue{Furthermore, by adapting existing backdoor methods for traditional deep neural networks, we propose a backdoor-based watermarking scheme (dubbed `BWAP') that embeds misclassification behaviors toward specific verification classes into soft prompts as shown in Figure \ref{comparison} (a).} However, we reveal that BWAP suffers from two crucial limitations: \emph{harmfulness} and \emph{ambiguity}\footnote{Different from \cite{fan2019rethinking}, the `ambiguity' is not defined as generating the potential watermark pattern of a given watermarked model. Instead, we discuss how to falsely claim the ownership of an independent object.}. \blue{The former indicates that BWAP introduces new security threats to the model, leading to misclassifying certain (verification) samples} (\ie, those containing defender-specified trigger patterns); the latter denotes that malicious developers can easily `fake' a watermark of independent soft prompts, leading to false claims of prompt ownership. In particular, we reveal that the reason for all these problems is that backdoor-based watermarking methods \emph{share the same decision space as the primary task}, yet their goal (\ie, to produce a specific misclassification as a distinctive behavior) is opposite to that of the primary task (\ie, correct classification). This central conflict explains the observed limitations. For the failure of existing CLIP backdoor methods, the conflict requires a large number of parameter updates and a significant training data budget to create a misclassifying association robust enough to override correct classification, which the prompt learning paradigm lacks. As for BWAP's limitations, its harmfulness is a direct consequence of the watermark's inherent conflict with the model's primary objective. The failure and harmfulness issues discussed above are direct results of the watermark's nature being opposite to the model's primary objective, whereas the last limitation, ambiguity, stems from the intrinsic constraint of the decision space itself. Specifically, both the primary task and the watermark task share the same low-complexity, binary decision space. This structural consistency forces the backdoor-based watermark to be simple, which inherently makes it easier to forge and more difficult to uniquely identify, thereby compromising both the security and reliability of the verification process.

In this paper, motivated by the insights above, we argue that a new watermarking paradigm is needed to overcome those limitations, one where the watermark's embedding space is totally different from the main task's decision space. To this end, we propose a new method called sequential watermarking for soft prompts (SWAP) to implant watermark information into a more complex space, as shown in Figure \ref{comparison} (b). In general, SWAP uses a particular order of a sequence of defender-specified out-of-distribution classes (dubbed `verification classes') as the watermark information. If the order of the predicted probability values of the suspicious model on these categories is the same as the defender-specified one, the model is regarded as containing the protected soft prompts. Our SWAP is inspired by the \emph{zero-shot} prediction characteristic of CLIP, whose prediction is performed via cosine similarities between image features (from the visual encoder) and textual features (from the text encoder) of any user-specified classes. In particular, our SWAP method introduces a novel watermarking task that is fundamentally different from traditional backdoor methods. It alters only the probability distribution over defender-specified verification classes while keeping the original prediction label unchanged, thereby ensuring that the process remains entirely harmless. Furthermore, the designed verification task is less opposite to the primary task, thereby requiring significantly fewer samples and minimal parameter updates, making it particularly suitable for prompt-learning settings. Furthermore, the probability ordering space across verification classes exhibits substantially higher complexity compared to the binary decision space, thereby increasing the difficulty of watermark counterfeiting and reducing ambiguity.

% In conclusion, our main contributions are four-fold. \textbf{(1)} We explore and formulate the copyright protection of public soft prompts as a specific type of model ownership verification problem. \textbf{(2)} We design a backdoor-based watermarking scheme (\ie, BWAP) for soft prompts and reveal its potential limitations. \textbf{(3)} We analyze the intrinsic reason for BWAP's limitations, based on which we propose a simple yet effective watermarking method (\ie, SWAP). SWAP is harmless and significantly reduces the risks of potential ambiguity attacks. \textbf{(4)} We conduct extensive experiments on eleven benchmark datasets to verify the effectiveness of our method and its resistance to potential adaptive attacks.

In conclusion, our main contributions are four-fold. \textbf{(1)} We explore and formalize the copyright protection of publicly released soft prompts as a specific form of model ownership auditing. \textbf{(2)} We systematically revisit existing non-intrusive auditing methods and intrusive backdoor-based watermarking approaches for CLIP, and reveal why these methods fail to effectively protect the copyright of soft prompts. Building on these insights, we extend backdoor watermarking techniques originally developed for traditional DNN classifiers and design the first effective backdoor-based watermarking framework for soft prompts, termed BWAP. \textbf{(3)} We further revisit backdoor-based watermarking methods, including BWAP, and identify that they not only suffer from potential ineffectiveness but also exhibit two fundamental limitations (\ie, ambiguity and harmfulness), which stem from the same underlying cause. Based on this analysis, we propose a simple yet effective watermarking method (dubbed SWAP), which is entirely harmless and substantially mitigates ambiguity risks. \blue{\textbf{(4)} We conduct extensive experiments on eleven datasets to verify the effectiveness and harmlessness of our method, and its resistance to potential adaptive attacks.}

\section{Related Work}\label{related_work}

\subsection{Vision Language Models and Prompt Tuning}
Vision-language models (VLMs) \citep{zhai2022lit, jia2021scaling, yu2022coca, radford2021learning} have made significant progress in learning joint representations of visual and textual information. CLIP \citep{radford2021learning} stands out as the pioneering approach that revolutionized this field by demonstrating remarkable zero-shot capabilities through contrastive learning on large-scale image-text pairs. Although these pre-trained models learn generalized representations, efficiently adapting them to downstream tasks while balancing computational cost, data efficiency, and performance remains challenging, particularly in resource-constrained scenarios.

Recent works \citep{zhou2022learning, zhou2022conditional,khattak2023self,xu2025progressive} have explored prompt tuning methods to efficiently adapt vision-language models like CLIP to downstream tasks without modifying the pre-trained weights. Different from traditional fine-tuning that updates all model parameters, prompt tuning only optimizes a small set of task-specific continuous vectors while keeping the foundation model frozen. CoOp \citep{zhou2022learning} pioneered this direction by introducing learnable continuous prompt optimization from downstream data. CoCoOp \citep{zhou2022conditional} further enhanced this approach by learning image-conditional prompts to improve model generalization across different domains. PromptSRC \citep{khattak2023self} also leveraged dual-modal prompting but adopted independent learnable prompts for text and image modalities, incorporating a self-regulating mechanism to acquire more task-agnostic knowledge. \blue{More recently, ATPrompt \citep{li2024advancing} advanced prompt learning by expanding the learning space of soft prompts from the category level to a multi-dimensional attribute level, anchoring universal attribute tokens into learnable prompts to achieve stronger generalization to unseen categories.} In particular, these learned soft prompts can be used as plug-and-play components to generate unique outputs that are distinct from the original CLIP model, making them valuable intellectual property assets.
%with 这些soft prompt（描述有一些特殊的输出）
%侵入主要是backdoor
%他因为是emb所以不比
\subsection{Model Ownership Auditing}
Model ownership auditing has emerged as a crucial technique for auditing models\footnote{We notice that dataset ownership auditing \cite{du2025sok,shao2025databench,li2025reliable} can also be used to protect the copyright of models. However, it is usually far less effective than model ownership auditing \cite{li2023black} since it can only modify training samples. As such, it is out of the scope of this paper.}. In general, this approach identifies unique passive or active signatures specific to the model, based on which to conduct ownership verification. Currently, almost all existing methods can be classified into two main categories: non-intrusive auditing methods and intrusive auditing methods. 

Non-intrusive auditing \citep{maini2021dataset,peng2022fingerprinting,dziedzic2022dataset} exploits the characteristics of a model's internal features after training to identify ownership without altering the model itself. Dataset Inference (DI) \citep{maini2021dataset} captures the decision boundaries by using the distance to each class to represent the learned features and trains a meta-classifier to verify whether a suspicious model possesses the same learned knowledge. Follow-up works focus on enhancing the representation of the learned inherent features. UAP \citep{peng2022fingerprinting} uses universal adversarial perturbations to represent the model's decision boundaries. By progressively adjusting the step size of a series of fixed-length trajectories, ADV-TRA \citep{adv-tra} creates an adversarial trajectory to represent internal features, thus overcoming the fragility of existing single-point methods against decision boundary changes. However, relying on learned features alone can easily lead to false positives, as different models may independently learn similar inherent features when trained on similar data distributions \citep{li2025move}. This vulnerability is especially pronounced for CLIP models in a few-shot setting, as their large-scale features remain unchanged while only a minimal number of parameters are modified, as shown in the following experiments in Section \ref{finger_limi}. As such, this paper mainly focuses on the intrusive auditing methods.

Intrusive auditing methods \citep{gu2017badnets,nguyen2021wanet,yang2023data,bai2024badclip,liang2024badclip} embed a unique signature directly into a model by introducing external or artificial features, creating an identifiable behavior that is triggered in the protected model but is rarely observed in others. \blue{Misclassification on specific samples (\ie, backdoor behaviors) is widely used as a type of intrusive auditing method \citep{shao2025sok}. }Numerous studies have employed backdoor-based watermarking methods \citep{ya2023towards,gan2023towards}, such as BadNet \citep{gu2017badnets} and WaNet \citep{nguyen2021wanet}, for the purpose of model auditing. Several pioneering backdoor attacks have also been proposed specifically for Vision-Language Models (VLMs), particularly targeting CLIP. BadEncoder \citep{jia2022badencoder} pioneered this direction by proposing the first backdoor attack on self-supervised learning, injecting backdoors into a pre-trained image encoder to ensure downstream classifiers inherit the malicious behavior. MmPoison \citep{yang2023data} introduced a data poisoning attack that is the first to study the vulnerability of multimodal models to attacks on both their visual and linguistic modalities. BadCLIP \citep{liang2024badclip} introduced a data poisoning attack that works by both ensuring visual triggers approximate textual semantics in the embedding space and aligning them with target vision features. Recent work \citep{tang2023watermarking,gao2025agate} has explored backdoor-based embedding watermarking methods for multi-modal Embedding-as-Service (EaaS) based on CLIP-based VLPs. Since these methods are embedding-based approaches, they are a type of white-box approach and thus fall outside the scope of this paper. \blue{To this end, research on model ownership auditing for soft prompts on CLIP models remains a blank and is worth exploration.}

% The former verifies model ownership by leveraging the inherent features that the model learned from training datasets, while the latter embeds unique and external distinctive prediction behaviors as the model watermark for ownership verification. The former is naturally harmless because no model modifications are required, but tends to false-positive judgments \cite{li2025move}. As such, this paper mainly focus on the watermarking-based MOV. Specifically, backdoor-based watermarking methods have become the mainstream approach in black-box scenarios. In general, backdoor-based watermarking methods embed distinctive response patterns that are activated by specific verification inputs. Detailed related works about backdoor attack are provided in the Appendix \ref{back_rela}. There are also a few pioneering studies exploring MOV for other tasks \cite{fernandez2023stable,wang2025sleepermark}. However, research on MOV for soft prompts in CLIP models remains blank and is worth further exploration.

\begin{figure*}
%\vspace{-1em}
  \centering
    \includegraphics[width=1\linewidth]{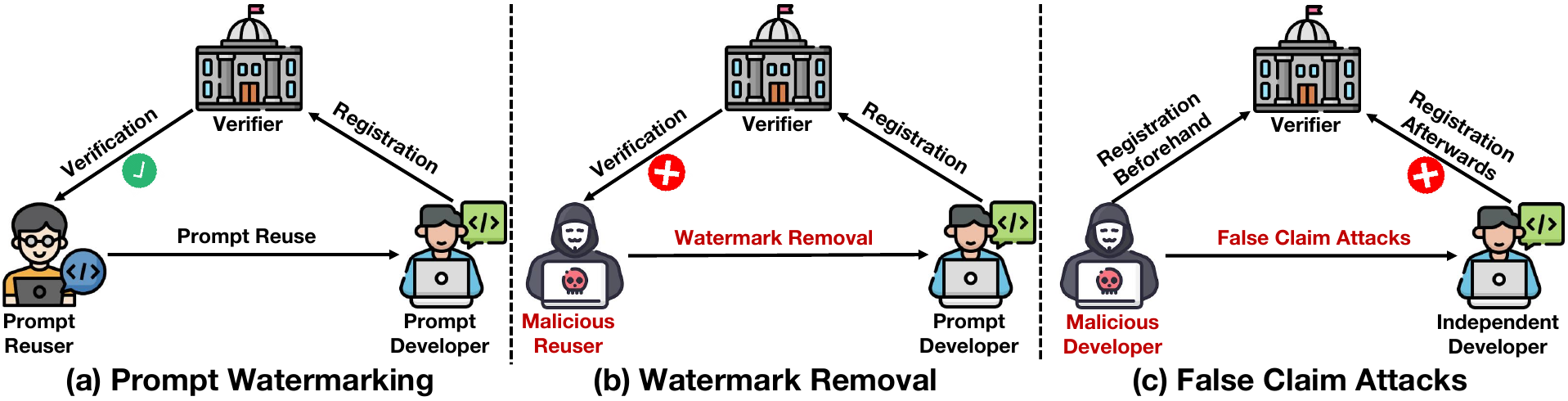}
%\vspace{0.3em}
  \caption{Three scenarios (one benign and two malicious) involved in prompt ownership verification. In prompt watermarking, the prompt developer generates a prompt along with its watermark and registers both with a trusted third-party verifier. When the prompt is maliciously reused, the verifier can determine its ownership by comparing the embedded watermark. In watermark removal attacks, a malicious reuser attempts to utilize the prompt while removing or obfuscating the watermark to evade verification, thereby enabling unauthorized use. In false claim attacks, a malicious developer seeks to pre-register a transferable watermark to falsely claim ownership of independently developed soft prompts for CLIP models.}
  % \vspace{-1em}
  \label{threat_model}
\end{figure*}

\section{Revisiting Existing Verification Method}\label{revisiting}
\subsection{Preliminaries}
\label{preliminaries}

\noindent \textbf{The Main Pipeline of Backdoor-based Watermarks in Image Classification.} \blue{Let $D = \{(\bm{x}_i, \bm{y}_i)\}_{i=1}^N$ denote the training set where $\bm{x}_i \in X = \{0, 1, ..., 255\}^{C \times W \times H}$ and $y_i \in Y = \{0, 1, ..., K-1\}$, and $K$ is the number of classes.} In the watermarking stage, the model owner selects $\gamma\%$ samples (\ie, $D_s$) to generate their poisoned version $D_p = \{(G_x(\bm{x}), G_y(y))|(\bm{x}, y) \in D_s\}$, where $G_x$ is the poisoned generator and $G_y$ is the label generator. For example, $G_x(\bm{x}) = (1 - \alpha) \oplus \bm{x} + \alpha \oplus \bm{t}$ and $G_y(y) = y_t$ in BadNets \citep{gu2017badnets}, where $\bm{t}$ is the selected trigger pattern, $\alpha$ is the trigger mask, and $y_t$ is the target label. \blue{The optimization objective for training a backdoor-watermarked model is $\min_{\theta} \sum_{({D} \setminus {D}_s) \cup {D}_p} L(f_{\theta}(\bm{x}_i), y_i)$, where $f_{\theta}$ is the model and $L$ is a given loss function.} In the verification stage, for a suspicious model $S$, the verifier uses watermarked samples $G_x(\bm{x})$ as verification samples to examine whether $S(G_x(\bm{x})) = G_y(y)$.

\begin{table*}[!t]
    \centering
    %\vspace{-1em}
    \caption{Results of three representative non-intrusive auditing verification methods for soft prompts trained on data with similar distributions. The low p-value, high cumulative distribution function (CDF) of similarities, and high fingerprint detection rate (FDR) indicate a false positive case, in which one independently trained soft prompt is incorrectly identified as a pirated copy of the other.}
% \vspace{-0.8em}
    % \resizebox{0.45\textwidth}{!}{
     \setlength{\tabcolsep}{3pt}
        \begin{tabular}{c|cccccc}
        \toprule
         Methods$\rightarrow$& \multicolumn{2}{c}{Dataset Inference \citep{maini2021dataset}}  & \multicolumn{2}{c}{UAP \citep{peng2022fingerprinting}}& \multicolumn{2}{c}{ADV-TRA \citep{adv-tra}}        \\
         \cmidrule(lr){2-3}\cmidrule(lr){4-5}\cmidrule(lr){6-7}
         Soft Prompts$\downarrow$& Accuracy& p-value& Accuracy& sim CDF($\uparrow$)& Accuracy& FDR($\uparrow$)\\
       \midrule
       MaPLe& 97.4\%& $10^{-4}$& 97.4\%&100\%& 97.4\%&100\%\\
       \midrule
       PromptSRC& 98.3\%& $10^{-5}$& 98.3\%&100\%& 98.3\%&100\%\\
       
        \bottomrule
    \end{tabular}
    % }
    
    % \vspace{-1em}
    \label{finger}
\end{table*}

\vspace{0.3em}
\noindent \textbf{Threat Model.} 
In this work, we aim to protect ownership of the soft prompts in CLIP models deployed as cloud services. Specifically, we consider three parties, including prompt developers, adversaries, and third-party verifiers. 

\begin{itemize}
  \setlength{\leftskip}{0pt}
    \item \noindent \textbf{Defenders' Goals and Capacities}. In the context of protecting the copyright of soft prompts, the defender corresponds to the prompt developer, as illustrated in Figure~\ref{threat_model}(a). Prior to releasing the prompt, the developer has full control over the training process, including access to proprietary training data, customized architectures, and the design of specific verification classes used for watermark detection. Once potential infringement is suspected, the verification process is conducted under a black-box setting, where the verifier has no access to the training data, architecture, or parameters of the suspicious model. The verifier can only interact with the suspicious model through its API interface, by querying it with the designed verification inputs and collecting the corresponding prediction probabilities.
    \item \noindent \textbf{Adversaries' Goals and Capacities}. We hereby consider two types of adversaries. As shown in Figure \ref{threat_model} (b), the first is malicious reusers who intend to acquire prompts by copying or stealing. In particular, they attempt to remove potential watermarks, operating under limited computational resources and data constraints. They may launch various watermark-removal attacks, such as fine-tuning and model pruning attacks, to compromise the watermark protection; As shown in Figure \ref{threat_model} (c), the second adversary is the model developer, who, operating in a complete black-box setting without knowledge of the soft prompt or training data, attempts to falsely claim the ownership of an independently developed soft prompt through the false claim attack. The formal definition of the false claim attack is as follows.
\end{itemize}

\begin{definition}
\blue{A false claim attack is an attempt by a malicious developer to falsely assert ownership of an independent soft prompt $\mathbf{V}_I$.} This is achieved by registering a set of fraudulent testing samples, $\mathcal{X}_{f}$, carefully crafted to pass the ownership verification. Given a CLIP model $\mathcal{M}$ and a ground-truth function $g(\mathbf{x})$, the attack is successful if for any testing sample $\mathbf{x} \in \mathcal{X}_{f}$, the condition $\mathcal{M}(\mathbf{V}_o, \mathbf{x}) = \mathcal{M}(\mathbf{V}_I, \mathbf{x}) \neq g(\mathbf{x})$ holds. Consequently, the verifier falsely asserts that $\mathbf{V}_I$ is a reused version of the adversary's soft prompt $\mathbf{V}_o$.
\end{definition}

%定义参考benchmark，不要一样

\subsection{The Limitation of Non-intrusive Auditing Methods}
\label{finger_limi}

Non-intrusive auditing methods verify model ownership by assessing whether a suspicious model encodes the same knowledge learned from a victim’s training data. This paradigm can also be extended to soft prompt ownership verification. However, such methods are prone to misjudgments when the suspicious model has been trained on data with a distribution similar to that of the victim. In these cases, independently trained models may capture analogous inherent features, leading the verification process to falsely identify a legitimately independent model as a pirated copy. The problem is further amplified in CLIP soft prompts, as their minor parameter adjustments preserve most of the model’s original representations.

\vspace{0.3em}
\noindent \textbf{Settings.} We generalize three representative non-intrusive auditing methods, including Dataset Inference~\citep{maini2021dataset}, UAP~\citep{peng2022fingerprinting}, and ADV-TRA~\citep{adv-tra}, to systematically evaluate this limitation. \blue{We conduct our experiments on the Caltech101 dataset \citep{fei2004learning} using two representative soft prompt methods (\ie, MaPLe \citep{khattak2023maple} and PromptSRC).} These two methods are alternately used as the independent model and the victim model for verification. As prompt learning is typically trained under a few-shot setting, we use different random seeds to sample two non-overlapping training subsets and independently train a soft prompt on each. This setup ensures that both soft prompts are derived from subsets with nearly identical data distributions yet remain independent. 
% \red{xxxx}

\vspace{0.3em}
\noindent \textbf{Results.} As shown in Table~\ref{finger}, the low p-value, high cumulative distribution function (CDF) of similarities, and high fingerprint detection rate (FDR) on the corresponding methods (Dataset Inference, UAP, and ADV-TRA) consistently and incorrectly identified one soft prompt as a pirated copy of the other, a conclusion that is demonstrably false. The extreme values of these metrics (low p-value, high CDF/FDR) demonstrate that this vulnerability is exacerbated in the prompt learning setting. Consequently, when two independent soft prompts are trained on highly similar data distributions, non-intrusive auditing methods incorrectly identify them as pirated copies.

\begin{table*}[t!]
\centering
\captionsetup{labelfont={color=tablecolor}, textfont={color=tablecolor}}
\caption{Adaptation of CLIP-based backdoor attacks to the MaPLe prompt learning method across three datasets.}
\vspace{-0.3em}
\setlength{\tabcolsep}{4pt}

\begin{tabular}{>{\color{tablecolor}}c|>{\color{tablecolor}}c>{\color{tablecolor}}c|>{\color{tablecolor}}c>{\color{tablecolor}}c|>{\color{tablecolor}}c>{\color{tablecolor}}c}
\toprule

Dataset$\rightarrow$
& \multicolumn{2}{c|}{\blue{ImageNet}}
& \multicolumn{2}{c|}{\blue{Caltech101}}
& \multicolumn{2}{c}{\blue{OxfordPets}} \\
\cmidrule(lr){2-3} \cmidrule(lr){4-5} \cmidrule(lr){6-7}

Backdoor Methods
& BA & ASR
& BA & ASR
& BA & ASR \\
\midrule

BadEncoder 
& 0.18 & 0.00 
& 1.29 & 0.00 
& 4.91 & 0.00 \\

mmPoison 
& 76.86 & 0.00
& 97.03 & 0.00 
& 95.53 & 0.00 \\

BadCLIP-D 
& 77.67 & 0.22 
& 97.15 & 7.42 
& 96.17 & 3.98 \\

\bottomrule
\end{tabular}
\vspace{-1em}
\label{badclip}
\end{table*}

\subsection{Limitations of Backdoor-based Intrusive Auditing}
\label{limi_back}

Backdoor-based intrusive auditing methods constitute a cornerstone of black-box ownership verification for deep neural networks \citep{li2025move}. In this section, we identify and analyze the limitations of two potential watermarking approaches: first, backdoor attacks specifically designed for CLIP models \citep{liang2024badclip}; second, the direct extension of backdoor attacks developed for conventional classification models to the CLIP setting. For each approach, we discuss the architectural and threat-model mismatches that undermine attack effectiveness and the practical obstacles to reliable ownership verification.

\vspace{0.3em}
\noindent \textbf{The Limitation of Using Backdoor Attack on CLIP for Watermarking.} The primary limitation is that when traditional CLIP backdoor watermarking methods are directly adapted to the prompt learning paradigm, they fail to embed the misclassification behavior, rendering the watermarks ineffective. This failure stems from a fundamental difference in scale: the methods designed for the CLIP model architecture cannot function within the parameter-efficient module of soft prompts. Fundamentally, we argue that the underlying reason for these failures is that the watermark shares the same decision space as the primary task, yet its goal is the complete opposite of the model's objective. This objective opposition requires a large number of parameter updates and a significant training data budget to create a `misclassifying association' significant enough to override correct classification, which the prompt learning paradigm lacks. 

To verify it, we adapt three backdoor attacks designed for the CLIP models, including BadEncoder \citep{jia2022badencoder}, mmPoison \citep{yang2023data}, and BadCLIP \citep{liang2024badclip}, to the CLIP prompt learning framework. \blue{Specifically, we adapt these attacks to a prompt-learning setting using the MaPLe \citep{khattak2023maple} on Caltech101 \cite{fei2004learning}, ImageNet \cite{deng2009imagenet}, and OxFordPets \cite{parkhi2012cats}. For the experiments, we strictly follow the original settings of each method. As shown in Table~\ref{badclip}, our results indicate that this adaptation proves ineffective, as all three backdoor attacks exhibit a very low attack success rate (ASR) when applied to prompt learning. These results suggest that backdoor attacks designed for CLIP cannot directly adapt to watermark CLIP's soft prompts. The experiment is intended as a preliminary motivation to illustrate why existing CLIP-specific backdoor attacks fail when naively adapted to prompt learning. To further validate the generality of this claim, we present comprehensive results in Table~\ref{main_res}, which covers diverse datasets and representative prompt learning methods.}

\begin{figure*}[t!]
    \centering
    \begin{tabular}{@{\extracolsep{\fill}}c@{}c@{\extracolsep{\fill}}}
            \includegraphics[width=0.473\linewidth]{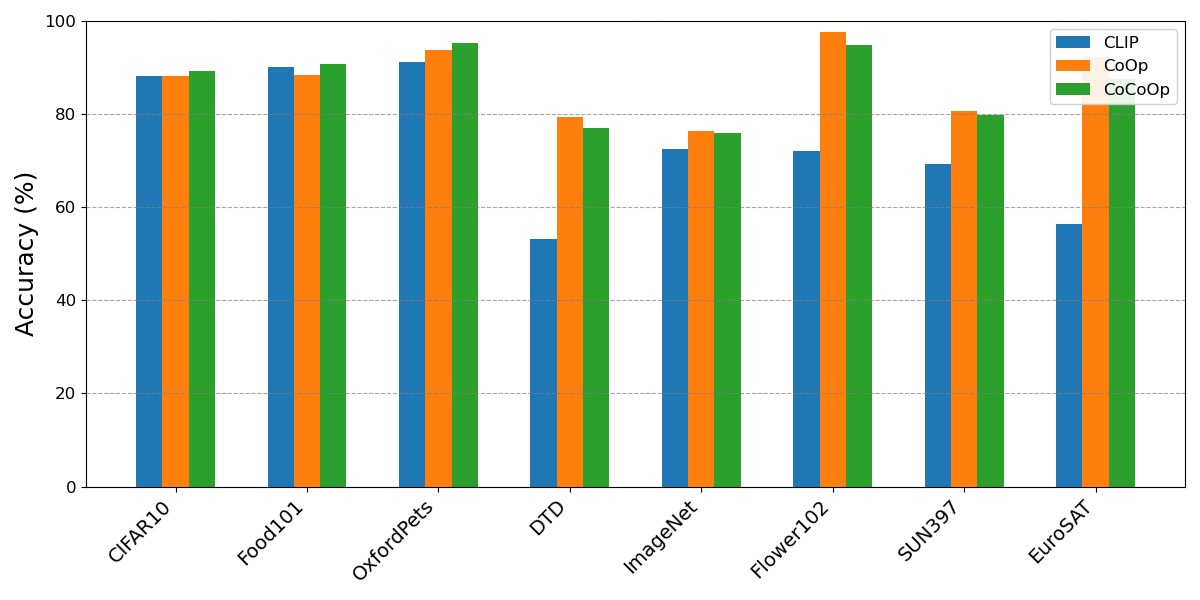} &
            \hspace{0.5em} \includegraphics[width=0.473\linewidth]{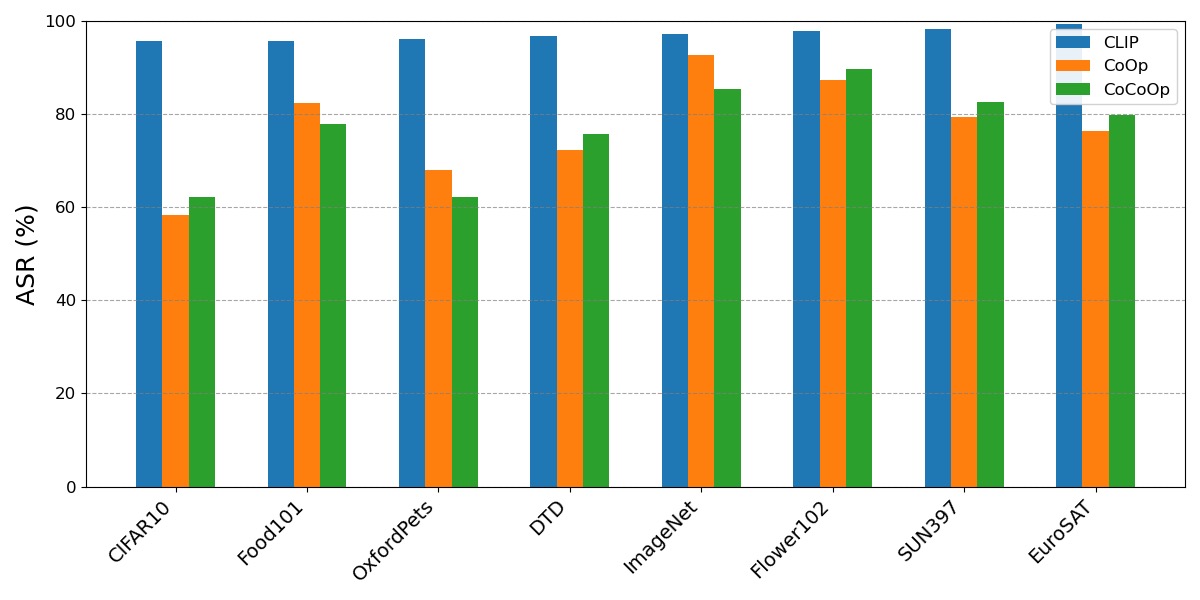}\\
            (a) Benign Accuracy & (b)Attack Success Rate \\
    \end{tabular}
    % \vspace{-2mm}
    \caption{\textbf{(a)} Benign accuracy on CoOp and CoCoOp across different datasets, showing the effectiveness of prompt tuning; \textbf{(b)} Attack success rate on CLIP (reference model), CoOp and CoCoOp (victim model), where adversarial examples generated by CLIP achieve high success rate on CLIP, CoOp and CoCoOp. The high transferability of attacks indicates that BWAP are susceptible to false claim attacks.}
    \vspace{-1em}
    \label{adv_trans_backdoor}
 \end{figure*}

\vspace{0.3em}
\label{limitation_bwap}
\noindent 
\textbf{Limitations of Backdoor-based Watermarking for Soft Prompts (BWAP).} 
Since direct application of existing backdoor attacks against CLIP models fails to embed the watermark, we further explore how to adapt existing backdoor methods for traditional deep neural networks and hereby propose BWAP (\textbf{b}ackdoor-based \textbf{wat}ermarking for soft \textbf{p}rompts). Specifically, we perform the embedding during the prompt tuning phase. With the CLIP image encoder $f(\cdot)$ and text encoder $g(\cdot)$ frozen, $\bm{V}_{f}$ and $\bm{V}_{g}$ are the tunable prompts. The prompt developer designates a specific target class $t$. We aim to achieve two objectives by updating $\bm{V}_{f}$ and $\bm{V}_{g}$, including maintaining correct classification on clean images x and embedding a backdoor watermark that causes triggered images $G(\bm{x})$ to be misclassified to the specific target class $t$ as mentioned in Section~\ref{preliminaries}. During verification stage, the verifier examines the existence of the backdoor in a suspicious model $S(\cdot)$ by checking if triggered images are classified to the target class, \ie, $S(G(\bm{x})) = t$. To increase the verification confidence and minimize the impact of randomness, we design a hypothesis test-based method following the previous work\citep{li2023black}. More details about the hypothesis test are in the Appendix $\ref{set_bwap}$.

% For any input image $\bm{x}$, the prediction probability for class $i$ is $p_i(\bm{x}) = \frac{\exp(sim(f(\bm{V}_f,\bm{x}), g(\bm{V}_g,\bm{c}_i))/\tau)}{\sum_{j=1}^{K+1} \exp(sim(f(\bm{V}_f,\bm{x}), g(\bm{V}_g,\bm{c}_j))/\tau)},$ where $sim(\cdot)$ is cosine similarity and $\tau$ is the temperature. 

Despite the effectiveness of BWAP, it still suffers from two critical limitations. The first is the \emph{harmfulness}. Backdoor-based watermark embeds patterns that trigger misclassification. While preserving performance on clean samples, they introduce new security vulnerabilities that adversaries could exploit defender-specified trigger pattern to misclassify verification samples. The second limitation is the \emph{ambiguity}, where malicious developers are able to fake a watermark that can falsely claim ownership of independent prompt. Specifically, the adversaries will generate adversarial examples $\bar{\bm{x}}$ based on the source model $M_o$ such that both the source model $M_o$ and the independent model $M_I$ will misclassify $\bar{\bm{x}}$ into the same incorrect class, \ie, $M_o(\bar{\bm{x}}) = M_I(\bar{\bm{x}})$. For example, $\bar{\bm{x}}$ can be generated through PGD attack\footnote{We hereby use the untargeted attack setting to enhance transferability, as targeted attacks may lead to overfitting to the source model. Notably, most adversarial examples generated this way are misclassified into the same incorrect class on source and independent models.}  that
\begin{equation}
    \bar{\bm{x}}_{t+1} =\mathit{Clip}_{\bm{x},\epsilon }(\bar{\bm{x}}_t + \gamma \cdot sign(\nabla\mathcal{J}(M_o ;\bar{\bm{x}}_t,\bm{y})),
\end{equation}
where $\mathit{Clip}_{\bm{x},\epsilon}(\cdot)$ constrains the perturbation within $\epsilon$ under the $\ell_\infty$ norm, $\sign(\cdot)$ is the sign function, $J(\cdot)$ is the
loss function associated with the original task of $M_o$ (\ie, CE loss in CLIP), and $\gamma$ is the step size. As shown in Figure \ref{adv_trans_backdoor}, even if we use PGD attack ($\epsilon$ = 8 / 255) instead of those with better transferability, adversarial examples generated on a reference model (\ie, CLIP) maintain high ASR on independent prompt-tuned models (\ie, CoOp and CoCoOp), demonstrating the transferability of these examples. 

We argue that both limitations are also rooted in the inherent conflict of objectives that the watermark shares the same decision space as the primary task, yet its goal is the opposite of the model's objective. The harmfulness is a direct result of the watermark's inherent conflict with the primary objective, while ambiguity stems from the intrinsic constraint of the decision space. The shared low-complexity, binary structure makes the signature inherently easier to forge.

\begin{figure*}
  \centering
  % \vspace{-1em}
    \includegraphics[width=\linewidth]{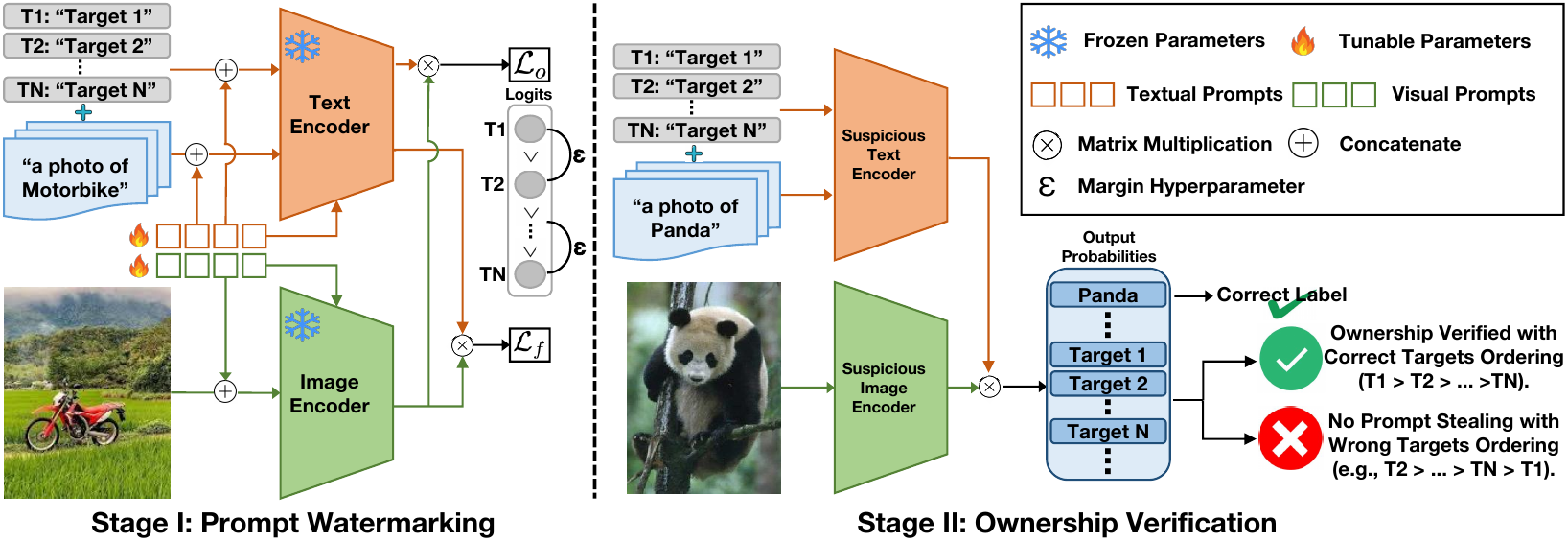}
  %\vspace{0.3em}
  \caption{The overall pipeline of SWAP. In the prompt watermarking stage, we introduce $\mathit{n}$ additional verification classes and embed a distinctive watermark by enforcing a predefined sequential ordering among them during prompt tuning. Specifically, the objective $\mathcal{L}_{o}$ applies a hinge-like constraint that maintains a fixed margin $\varepsilon$ between the logits of consecutive verification classes, while $\mathcal{L}_{f}$ preserves the original task performance through a standard cross-entropy loss. In the ownership verification stage, we incorporate the verification classes into the testing classes and examine the predetermined sequential pattern in the output probabilities to verify the ownership.
  %The main pipeline of our SWAP. In the prompt watermarking stage, we specify $\mathit{n}$ additional verification classes and embed a distinctive watermark by establishing a predetermined sequential ordering of the verification classes during prompt tuning. Specifically, $\mathcal{L}_{o}$ implements a hinge-like constraint to establish a fixed margin $\varepsilon$ between the logit values of consecutive verification classes, while $\mathcal{L}_{f}$ preserves the prompt's classification performance through CE loss. In the ownership verification stage, we incorporate the verification classes into the testing classes and examine the predetermined sequential pattern in the output probabilities to verify the ownership.
  }
  \label{pipeline}
%\vspace{-1em}
\end{figure*}

%三个问题的原因是一个，所以设计一个针对这个原因的方法

\section{The Propose Method}\label{sec4}

Motivated by the above insights, we argue that overcoming the intrinsic limitations of existing approaches requires a new watermarking paradigm, which decouples the watermark’s embedding space from the model's primary decision space. To this end, we introduce Sequential Watermarking for Soft Prompts (SWAP), a novel framework that embeds watermark information into a more complex space by exploiting the zero-shot prediction capability of CLIP models.

In general, our SWAP consists of two stages: prompt watermarking and ownership verification as shown in Figure~\ref{pipeline}. In general, it uses an owner-specified order of a sequence of defender-specified additional verification classes as the watermark. It implants the sequential watermark into the protected soft prompt in the first stage and then verifies whether a third-party suspicious model infringes our prompts based on its prediction of verification classes in the second stage. 

% This novel task fundamentally differs from traditional backdoors by only altering the probability distribution over verification classes while keeping the original label unchanged, ensuring the process is harmless.

% \red{Motivated by previous understandings, We propose a new paradigm that avoids the core conflict between the watermark task and the main classification task by using a totally different space.Our method, sequential watermarking for soft prompts (SWAP), implant the watermark information into a more complicated space by leveraging the zero-shot prediction characteristic of CLIP models. In general, SWAP exploits an owner-specified order of a sequence of given classes (dubbed `verification classes') as the watermark.}

% \red{Specifically, our SWAP consists of two main stages: prompt watermarking and ownership verification. It implants the sequential watermark into the protected soft prompt in the first stage and verifies whether a third-party suspicious model infringes our prompts based on its prediction of verification classes, as shown in Figure \ref{pipeline}.}

\subsection{Prompt Watermarking}

In general, the watermarking process can be formulated as a multi-objective optimization problem: \textbf{(1)} maintaining the model's original classification performance on all samples, and \textbf{(2)} embedding a predefined sequential ordering of prediction probabilities across verification classes to serve as the watermark. Let $\mathcal{T} = \{t_i\}_{i=1}^n$ denotes our selected verification classes for prompt watermarking (\eg,``Target 1",..., ``Target $\{n\}$") while $\{\bm{c}_{t_i}\}_{i=1}^n$ denotes their corresponding word embeddings. Similarly, let $\{\bm{c}_i\}_{i=1}^K$ denote the word embeddings of the original classes. In this case, for any input image $\bm{x}$, the prediction probability for class $i$ is $p_i(\bm{x}) = \frac{\exp(sim(f(\bm{V}_f,\bm{x}), g(\bm{V}_g,\bm{c}_i))/\tau)}{\sum_{j=1}^{K} \exp(sim(f(\bm{V}_f,\bm{x}), g(\bm{V}_g,\bm{c}_j))/\tau)}$ as well. As such, the optimization objective can be formalized as: 
\begin{equation}
\min_{\theta_{\bm{V}_f},\theta_{\bm{V}_g}} \mathcal{L} = \mathcal{L}_{f} + \lambda \cdot \mathcal{L}_{o},
\label{overall_loss}
\end{equation}
where $\lambda$ is a trade-off hyper-parameter and $\theta_{\bm{V}_f}, \theta_{\bm{V}_g}$ are the parameters of tunable prompts $\bm{V}_f$ and $\bm{V}_g$, respectively.

The first term, \ie, functionality loss $\mathcal{L}_{f}$ is used to maintain the performance on the original classes, \ie, 
\begin{equation}
\mathcal{L}_{f} = -\sum_{\mathcal{D}} \bm{y}_i \log(p_i(\bm{x})).
\end{equation}

In the second term, \ie, order loss $\mathcal{L}_{o}$, we maintain equal intervals between the logits of verification classes designated by the prompt developer to establish a specific ordering that serves as the watermark. Specifically, let $\{z_i\}_{i=1}^n$ denote the logits of verification classes, where $z_i = sim(f(\bm{V}_f,\bm{x}), g(\bm{V}_g,\bm{c}_{t_i}))/\tau$ for each image $\bm{x}$. To achieve this, we design a hinge-like loss, as follows:

\begin{equation}
\mathcal{L}_{o} = \sum_{\mathcal{D}} \sum_{i=1}^{n-1} \max(0, \varepsilon - (z_{i+1} - z_i)),
\label{order}
\end{equation}
where $\varepsilon$ is a margin hyperparameter. This loss ensures that consecutive verification class logits maintain a margin of $\varepsilon$ (\ie, $ z_{i+1} - z_i \geq \varepsilon, \ \forall i \in [1,n-1].$ By optimizing the loss in Eq.~(\ref{overall_loss}), the watermark is embedded through ordered logits among verification classes while maintaining the model's performance on non-verification classes, no matter they are \emph{seen} or \emph{unseen}.

\subsection{Ownership Verification} 

In this section, we introduce how to conduct prompt ownership verification based on our SWAP. Given the suspicious model $S(\cdot)$, the defenders can use the verification sample $\bm{x}$ to obtain the probability sequence $p=\{S(\bm{x})_i\}_{i=t_1}^{t_n}$ of the defender-specified verification classes $\mathcal{T} = \{t_i\}_{i=1}^n$. 
Then they can verify potential prompt stolen by examining whether the order sequence $\pi(p)$ matches the predefined sequence $\pi_{o}(\mathcal{T})$ where $\pi(\cdot)$ is the sorting function. However, the verification result may be sharply affected by the randomness of selecting $\bm{x}$, In order to increase the verification confidence, we design a hypothesis test-guided method, as follows:

\begin{algorithm}[!t]
    \renewcommand{\algorithmicrequire}{\textbf{Input:}}
    \renewcommand{\algorithmicensure}{\textbf{Output:}}
  % \setstretch{1.2}
  \caption{Ownership verification based on pair-wise hypothesis test.}
  \label{algo:detail_frame}
  \begin{algorithmic}[1]
  \REQUIRE verification dataset $\mathcal{D} = \{(\bm{x}_i, y_i)\}_{i=1}^{m}$, suspicisous model $f$, target classes $\mathcal{T} = \{t_i\}_{i=1}^n$, original sequence $\pi_{o}(\mathcal{T})$, threshold $\tau$
  \ENSURE A boolean value indicating whether passing the ownership verification process.
    
    \STATE $p$ = extraction($\mathcal{D}$, $f$, $\mathcal{T}$)
    \STATE $\pi(p)$ = sort($p$)
    \STATE $p\text{-value}$ = T-TEST($d(\pi(p), \pi_{o}(\mathcal{T}))$, 0, $\tau$)
    \IF{$p\text{-value} \leq \alpha$}
      \STATE \textbf{return} \text{True}
    \ELSE
      \STATE \textbf{return} \text{False}
    \ENDIF
    \end{algorithmic}
    \label{alg_veri}
    
\end{algorithm}

\begin{proposition}
Let $\pi(p)$ be the sequence extracted from the suspicious model and $\pi_{o}(\mathcal{T})$ is the defender-specified sequence. Given the null hypothesis: $H_0: d(\pi(p), \pi_{o}(\mathcal{T})) = \tau $ and the alternative hypothesis: $
H_1: d(\pi(p), \pi_{o}(\mathcal{T})) < \tau$, where $\tau$ is a threshold parameter and $d$ represents the total distance between the extracted and the original sequence, we claim that the suspicious model is an unauthorized copy (with $\tau-certainty$) if and only if $H_0$ is rejected.
\end{proposition}
In practice, we randomly select $m$ samples to conduct the one-sided T-test \citep{larsen2005introduction} and calculate the p-value. If the p-value is less than a given significance level $\alpha$ (\eg, 0.01), the null hypothesis will be rejected and the suspicious model can be regarded as containing the protected soft prompt. We use the the sum of Absolute Rank Difference to calculate the distance $d$. The Ownership Verification Algorithm of our SWAP is presented in Algorithm \ref{alg_veri}. Specifically, we first extract the probabilities sequence of the verification classes from the suspicious model to construct the extracted sequence. We use T-test to compare the extracted sequence with the defender-specified sequence, returning true (successful verification) if the p-value is less than significance level $\alpha$.

% The pseudocode of the ownership verification algorithm is presented in Appendix \ref{algo_psudo}.

Having established the hypothesis test-based prompt ownership verification based on SWAP, we now theoretically analyze the success conditions of SWAP-based ownership verification as below.

\begin{theorem}
Let $S(\bm{x})$ be the posterior probability of $\bm{x}$ predicted by the suspicious model, variable $\bm{X}$ denotes the test sample with verification classes. When the extracted sequence differs from the original sequence, \blue{we assume that their distances follow a distribution over $\{2, 4, ..., \lfloor\frac{n^2}{2}\rfloor\}$.} In this case, we claim that verifiers can reject the null hypothesis $H_0$ at the significance level $\alpha$, if the average distance $d$ of $S$ satisfies that
\begin{equation}
0\le d<\frac{2(m - 1)\tau + t_{\alpha}^2 - \sqrt{\Delta}}{2\left[(m-1) + t_{\alpha}^2\right]},
\end{equation}
where $\Delta=a^2 t_{\alpha}^4 + 4(m-1) t_{\alpha}^2 \tau (a - \tau)>0$, $t_{\alpha}$ is the $(\alpha)$-quantile of t-distribution with $(m - 1)$ degrees of freedom, m is the sample size of $X$, and a serves as the upper bound of all possible values of $d$.
\label{theorem_1}
\end{theorem}

\blue{In general, Theorem \ref{theorem_1} serves as a theoretical analysis that provides a sufficient condition under which SWAP-based ownership verification can succeed. Specifically, it demonstrates that the average distance $d$ is not exactly zero but remains sufficiently small, which aligns with the statistical properties of hypothesis testing.} The detailed proof is in Appendix \ref{proof}. 

% In general, Theorem \ref{theorem_1} provides a theoretical analyze for SWAP, demonstrating that the SWAP-based ownership verification can succeed even if the average distance $d$ is not exactly zero but remains sufficiently small, which aligns with the statistical properties of hypothesis testing. The detailed proof is in Appendix \ref{proof}. 

\begin{table*}[t!]
%\vspace{-1em}
    \centering
    \caption{Watermarking and verification performance of SWAP compared with baseline methods. We highlight the superior results of each method in bold and mark results demonstrating a negative impact on model utility in \red{red}.}
    % \vspace{-0.8em}
    \resizebox{1\textwidth}{!}{
     \setlength{\tabcolsep}{3pt}
        \begin{tabular}{c|c|ccccc |ccccc |ccccc}
        \toprule
          & Dataset{$\rightarrow$}
         & \multicolumn{5}{c}{ImageNet}
         & \multicolumn{5}{c}{Caltech101} &
          \multicolumn{5}{c}{OxfordPets} 
         % &\multicolumn{5}{c}{Food101}
         \\
        
           \cmidrule(lr){3-7}  \cmidrule(lr){8-12} \cmidrule(lr){13-17} 
           %\cmidrule(lr){18-22}

        \makecell{Prompt Tuning \\Method$\downarrow$}& \makecell{Protection  \\Method$\downarrow$}
        & \makecell{ACC \\(Base)}    & \makecell{ACC \\(Novel)}  & WSR     &p-value &$\hat{H}$
        & \makecell{ACC \\(Base)}    & \makecell{ACC \\(Novel)}  & WSR     &p-value &$\hat{H}$
        & \makecell{ACC \\(Base)}    & \makecell{ACC \\(Novel)}  & WSR     &p-value &$\hat{H}$
        % & Base    & Novel  & WSR     &p-value &$\hat{H}$
        \\
       \midrule
        \multirow{8}{*}{\makecell{CoCoOp}} 
        % & Ind Prompt & 75.98 & 70.43 & 4.37 &  0.70 & - & 97.96 & 93.81 & 3.06 &  0.70 & - & 95.20 &97.69 & 6.43 &  0.70 & - 
        % % & 90.70& 91.29& 1.92&  0.70 & -
        % \\ 
        % & Ind Target & 75.98 & 70.43 & 4.37 &  0.70 & - & 99.98 & $10^{-4848}$ & 0.00 &  0.70 & - & 98.60 & 0.69 & 72.07 &  0.70 & - 
        % % & 0.70& 99.60 & 0.70&  0.70 & -
        % \\ 
        & BadCLIP-T & 75.60 & 70.00 & 99.90 & $10^{-80}$ & \red{0.70} & 98.00 & 93.00 & 99.20 & $10^{-37}$ & \red{0.93} & 92.60 & 95.70 & 99.20 & $10^{-23}$ & \red{0.95} 
        %& 88.90& 89.80 & 98.40& 99.60 & 0.90
        \\ 
        & BadEncoder & 4.21 & 1.70 & 4.63 & $10^{-1}$ & \red{-0.25} & 7.37 & 2.30 & 8.39 & $10^{-1}$ & 0.02 & 5.19 & 4.90 & 4.63 & $10^{-1}$ & 0.03 
        %& 88.90& 89.80 & 98.40& 99.60 & 0.90
        \\ 
        & mmPoison & 75.24 & 70.04 & 0.02 & $10^{-1}$ & \red{-0.30} & \textbf{98.06} & 93.01 & 0.00 & $10^{-1}$ & -0.06 & 94.37 & 97.07 & 0.00 & $10^{-1}$ & -0.02
        %& 88.90& 89.80 & 98.40& 99.60 & 0.90
        \\ 
        & BadCLIP-D & 75.59 & 70.05 & 0.02 & $10^{-1}$ & \red{-0.25} & 97.50 & 93.03 & 8.52 & $10^{-1}$ & 0.02 & 94.68 & 97.08 & 3.77 & $10^{-1}$ & 0.02 
        %& 88.90& 89.80 & 98.40& 99.60 & 0.90
        \\ 
        & BWAP-BadNet & 73.15 & 67.60 & 98.60 & $10^{-119}$ & \red{0.69} & 95.10 & 93.10 & 97.20 & $10^{-43}$ & \red{0.91} & 91.30 & 93.90 & 99.00 & $10^{-46}$ & \red{0.97} 
        %& 88.90& 89.80 & 98.40& 99.60 & 0.90
        \\ 
        & BWAP-WaNet & 72.07 & 67.60 & 99.60 & $10^{-120}$ & \red{0.70} & 95.20 & 90.80 & 97.90 & $10^{-46}$ & \red{0.92} & 92.30 & 94.20 & 98.90 & $10^{-41}$ & \red{0.97} 
        %& 88.70& 96.90 & 99.50 & 99.50 & 0.91
        \\ 
        & BWAP-Grond & 72.85 & 68.42 & 99.32 & $10^{-118}$ & \red{0.70} & 95.16 & 92.45 & 98.41 & $10^{-49}$ & \red{0.92} & 92.95 & 94.57 & 98.95 & $10^{-44}$ & \red{0.97} 
        %& 88.70& 96.90 & 99.50 & 99.50 & 0.91
        \\ 
        & \textbf{SWAP (ours)} & \textbf{75.89} & \textbf{70.10} & \textbf{99.98} & \textbf{0} & \textbf{0.00} & 97.65 & \textbf{93.10} & \textbf{99.67} & \textbf{0} & \textbf{0.01} & \textbf{94.92} & \textbf{97.40} & \textbf{99.36} & \textbf{0} & \textbf{0.00} 
        %& 90.30 & 91.18 & 99.79 & $10^{-368}$ & 0.00
        \\ 
        \midrule
        \multirow{7}{*}{MaPLe} 
        % & Ind Prompt & 76.66 &70.54& 5.46 &  0.70 & - & 97.74 &94.36& 2.51 &  0.70 & - & 97.69& 97.76& 3.24 &  0.70 & - 
        % %& 90.71& 92.05& 3.61 &  0.70 & -
        % \\ 
        % & Ind Target & 75.98 & 70.43 & 4.37 &  0.70 & - & 99.98 & $10^{-4848}$ & 0.00 &  0.70 & - & 98.60 & 0.69 & 72.07 &  0.70 & - 
        % %& 0.70& 99.60 & 0.70&  0.70 & -
        % \\ 
        & BadEncoder & 0.24 & 0.20 & 4.31 & $10^{-1}$ & \red{-0.25} & 8.84 & 1.10 & 0.00 & $10^{-1}$ & -0.02 & 5.21 & 5.40 & 2.13 & $10^{-1}$ & 0.00 
        %& 88.90& 89.80 & 98.40& 99.60 & 0.90
        \\ 
        & mmPoison & 76.24 & \textbf{69.80} & 0.00 & $10^{-1}$ & \red{-0.30} & 97.26 & 92.80 & 0.00 & $10^{-1}$ & -0.02 & 94.74 & \textbf{97.90} & 0.00 & $10^{-1}$ & -0.02
        %& 88.90& 89.80 & 98.40& 99.60 & 0.90
        \\ 
        & BadCLIP-D & 77.04 & 69.04 & 0.02 & $10^{-1}$ & \red{-0.30} & \textbf{98.00} & 93.90 & 8.65 & $10^{-1}$ & 0.07 & 94.96 & 96.04 & 3.99 & $10^{-1}$ & 0.02
        %& 88.90& 89.80 & 98.40& 99.60 & 0.90
        \\ 
        & BWAP-BadNet & 76.70& 66.80& 99.50 & $10^{-154}$ & \red{0.70} & 97.60 & 94.00& 99.20 & $10^{-71}$ & \red{0.94} & 94.50& 94.90 &98.90& $10^{-57}$ & \red{0.97} 
        %& 89.80& 90.60& 99.70 & 0.70& 0.92
        \\ 
        & BWAP-WaNet & 76.60& 69.30 & 99.70 & $10^{-173}$ & \red{0.70} & 97.50 &92.03& 99.60 & $10^{-97}$ & \red{0.98} & 94.80 &95.90& 99.90 & $10^{-49}$ & \red{0.98} 
        %& 89.50& 90.20& 99.90 & 0.70& 0.92
        \\ 
        & BWAP-Grond & 76.28 & 68.73 & 98.50 & $10^{-138}$ & \red{0.69} & 97.40 & 92.60 & 98.40 & $10^{-58}$ & \red{0.97} & 94.71 & 95.35 & 99.61 & $10^{-53}$ & \red{0.98} 
        %& 88.70& 96.90 & 99.50 & 99.50 & 0.91
        
        \\ 
        & \textbf{SWAP (ours)} & \textbf{77.13} & 69.26 & \textbf{99.95} & \textbf{0} & \textbf{0.01} & 97.30 & \textbf{95.31} & \textbf{99.99} & \textbf{0} & \textbf{0.01} & \textbf{95.03} & 96.82 & \textbf{99.94} & \textbf{0} & \textbf{0.01} 
        %& 90.32& 91.33 & 99.97 & $10^{-2939}$ & 0.01
        \\ 
        \midrule
        \multirow{7}{*}{PromptSRC} 
        % & Ind Prompt & 77.60& 70.73& 3.31 &  0.70 & - & 98.10 &94.03 &3.38 &  0.70 & - & 95.33& 97.30 &5.48 &  0.70 & - 
        % %& 90.67& 91.53& 4.10&  0.70 & -
        % \\ 
        % & Ind Target & 75.98 & 70.43 & 4.37 &  0.70 & - & 99.98 & $10^{-4848}$ & 0.00 &  0.70 & - & 98.60 & 0.69 & 72.07 &  0.70 & - 
        % %& 0.70& 99.60 & 0.70&  0.70 & -
        % \\ 
        & BadEncoder & 5.48 & 1.10 & 0.10 & $10^{-1}$ & \red{-0.30} & 7.17 & 3.60 & 3.16 & $10^{-1}$ & -0.03 & 4.57 & 6.80 & 29.03 & $10^{-1}$ & \red{0.26} 
        %& 88.90& 89.80 & 98.40& 99.60 & 0.90
        \\ 
        & mmPoison & 75.25 & 70.10 & 0.00 & $10^{-1}$ & \red{-0.30} & \textbf{98.19} & 94.04 & 0.00 & $10^{-1}$ & -0.06 & 95.16 & 97.50 & 0.00 & $10^{-1}$ & -0.03
        %& 88.90& 89.80 & 98.40& 99.60 & 0.90
        \\ 
        & BadCLIP-D & 77.26 & 70.07 & 0.00 & $10^{-1}$ & \red{-0.30} & 98.10 & 94.20 & 8.46 & $10^{-1}$ & 0.02 & 95.64 & \textbf{97.80} & 4.25 & $10^{-1}$ & 0.01 
        %& 88.90& 89.80 & 98.40& 99.60 & 0.90
        \\ 
        & BWAP-BadNet & 76.90& \textbf{70.80}& 99.60& $10^{-131}$ & \red{0.70} & 97.40& 93.30& 99.30 & $10^{-73}$ & \red{0.93} & 95.30& 97.10& 98.00 & $10^{-63}$ & \red{0.95} 
        %& 90.30& 91.10& 99.60 & 0.70& 0.91
        \\ 
        & BWAP-WaNet & 76.60 &70.40& 98.70 & $10^{-128}$ & \red{0.69} & 97.70 & 93.70& 99.20 & $10^{-69}$ & \red{0.93} & 95.50& 96.90& 98.70 & $10^{-69}$ & \red{0.96} 
        %& 90.10& 91.03& 98.90 & 0.70& 0.90
        \\ 
        & BWAP-Grond & 76.75 & 70.65 & 99.13 & $10^{-123}$ & \red{0.70} & 97.55 & 93.43 & 99.15 & $10^{-67}$ & \red{0.93} & 95.45 & 97.31 & 98.40 & $10^{-65}$ & \red{0.95} 
        %& 88.70& 96.90 & 99.50 & 99.50 & 0.91
        \\ 
        & \textbf{SWAP (ours)} & \textbf{77.43} & 70.48 & \textbf{99.97} & \textbf{0} & \textbf{0.00} & 96.28 & \textbf{94.21} & \textbf{99.94} & \textbf{0} & \textbf{0.00} & \textbf{96.13} & 96.80 & \textbf{99.99} & \textbf{0} & \textbf{0.01} 
        %& 90.60 & 91.35 & 99.97 & $10^{-3182}$ & 0.00
        \\ 

       \bottomrule
    \end{tabular}
    }
    
    \label{main_res}
    %\vspace{-0.5em}
\end{table*}

\section{Experiments}\label{sec5}

\noindent\textbf{Baselines and Datasets.} In our experiments, we apply prompt tuning on a pretrained ViT-B/16 CLIP \citep{radford2021learning} model using the following methods: CoCoOp \citep{zhou2022conditional}, MaPLe \citep{khattak2023maple}, and PromptSRC \citep{khattak2023self}. We also conduct experiments using SOTA prompt tuning method ATPrompt \citep{li2024advancing} with results provided in Appendix \ref{sota_main}. For baselines, we first adapt representative traditional backdoor attacks (\eg, BadNet \citep{gu2017badnets} and WaNet \citep{nguyen2021wanet}) to the CLIP prompt learning setting, corresponding to BWAP-BadNet and BWAP-WaNet, respectively. We also adapt the state-of-the-art DNN-based backdoor attack, Grond \citep{xu2025towards}, to this setting, which we name BWAP-Grond. For backdoor attacks specifically designed for CLIP, we follow the setting of \ref{limi_back} to adapt BadEncoder \citep{jia2022badencoder}, mmPoison \citep{yang2023data}, and BadCLIP-D \citep{liang2024badclip} to the prompt learning framework. Additionally, for backdoor attacks on CLIP prompt learning, we also compared against the BadCLIP-T \citep{bai2024badclip} method. It is important to note this method is only applicable to CoCoOp \citep{zhou2022conditional} and is not generalizable to other prompt learning methods. For prompt length and depth settings, we follow the original configurations of each method. For all training-related configurations and strategies, we also follow the original settings of each method. Following previous works on prompt learning \citep{zhou2022conditional,zhou2022learning}, we evaluate the performance of our method on a subset of 11 image classification benchmark datasets, which covers a wide range of recognition tasks. Specifically, we conduct main experiments on Caltech101 \citep{fei2004learning}, ImageNet \citep{deng2009imagenet}, and OxfordPets \citep{parkhi2012cats}. Results on other datasets are in Appendix \ref{other_data}.

%Having demonstrated the effectiveness of SWAP within a single dataset, we further show that SWAP maintains its effectiveness in more challenging scenarios, including  cross-dataset and domain generalization scenarios, without compromising the soft prompt's own generalizability.  We train the prompt on ImageNet and directly evaluate it on other datasets without any data-specific fine-tuning.

\vspace{0.3em}
\noindent\textbf{Evaluation Metrics.} Following previous works on prompt learning \citep{zhou2022conditional,zhou2022learning}, we aim to evaluate the generalizability across various classes. This process involves dividing the dataset into base (\ie, seen) and novel (\ie, unseen) classes (dubbed `base-to-novel scenario') and then training the model using a small number of samples from the base classes. We evaluate the model's accuracy on both base (ACC (Base)) and novel (ACC (Novel)) classes. For watermarking evaluation, we use watermarking success rate (WSR) to measure the model's accuracy on verification samples. Inspired by the definition in  \citep{guo2023domain}, we define the relative harmless degree $\hat{H} \triangleq \frac{1}{N}(\sum^{N}_{i=1}\mathbb{I}(\hat{f}(\hat{\bm{x}}_i)\ne y_i) - \sum^{N}_{i=1}\mathbb{I}(f(\hat{\bm{x}}_i)\ne y_i))$ as the metric for evaluating the level of harmlessness, where $\hat{\mathcal{D}}={(\hat{\bm{x}}_i,y_i)}_{i=1}^N$ denotes a set of verification samples, $f(\cdot)$ and $\hat{f}(\cdot)$ represent the original independent model and the watermarked model respectively, and $\mathbb{I}(\cdot)$ is the indicator function. For prompt verification, we also conduct statistical hypothesis testing and use p-value to verify the effectiveness of verification. To conduct an in-depth study, we evaluate our method in independent prompt and independent verification classes settings. In the first setting, we test on independent prompts using pre-defined verification classes. In the second setting, we test on watermarked prompts using randomly selected independent verification classes. In both settings, a reliable verification method ought to have a larger p-value.

\label{main_settings}

\begin{table*}[!t]
 \centering
 \caption{\textnormal{Accuracy on cross-dataset benchmark evaluation.}}
     % \vspace{-0.8em}
    \setlength{\tabcolsep}{2.5pt}
    % \scalebox{0.6}{
    \begin{tabular}{l c ccccccccccc}
    \toprule
    & \textbf{Source} & \multicolumn{11}{c}{\textbf{Target}} \\ \cmidrule(lr){2-2} \cmidrule(lr){3-13}
    & \rotatebox{90}{ImageNet} & \rotatebox{90}{Caltech101} & \rotatebox{90}{OxfordPets} & \rotatebox{90}{StanfordCars} & \rotatebox{90}{Flowers102} & \rotatebox{90}{Food101} & \rotatebox{90}{Aircraft} & \rotatebox{90}{SUN397} & \rotatebox{90}{DTD} & \rotatebox{90}{EuroSAT} & \rotatebox{90}{UCF101} & \rotatebox{90}{\emph{Average}} \\
    \midrule
    
    Co-CoOp & 70.90 & 91.93 & 89.02 & 64.78 & 79.41 & 83.96 & 20.73 & 63.70 & 41.73 & 45.64 & 66.21 & 65.27 \\
    MaPLe & 70.55 & 92.86 & 90.13 & 65.07 & 71.30 & 86.36 & 21.39 & 66.81 & 45.39 & 44.86 & 68.09 & 65.71 \\
    
 PromptSRC & 70.53 & 93.10 & 89.78 & 64.12 & 70.52 & 86.52 & 23.02 & 67.33 & 46.27 & 45.88 & 68.97 & 66.00 \\
    \bottomrule
    \end{tabular}
    % }
    \vspace{-2em}
       
    \label{tab:xd_acc} %\vspace{-1em}
\end{table*}

\begin{table*}[!t]
\centering
\caption{\textnormal{WSR on cross-dataset benchmark evaluation.} }
    % \vspace{-0.8em}
    \setlength{\tabcolsep}{2.5pt}
    \scalebox{0.91}{
    \begin{tabular}{l c ccccccccccc}
    \toprule
    & \textbf{Source} & \multicolumn{11}{c}{\textbf{Target}} \\ \cmidrule(lr){2-2} \cmidrule(lr){3-13}
    & \rotatebox{90}{ImageNet} & \rotatebox{90}{Caltech101} & \rotatebox{90}{OxfordPets} & \rotatebox{90}{StanfordCars} & \rotatebox{90}{Flowers102} & \rotatebox{90}{Food101} & \rotatebox{90}{Aircraft} & \rotatebox{90}{SUN397} & \rotatebox{90}{DTD} & \rotatebox{90}{EuroSAT} & \rotatebox{90}{UCF101} & \rotatebox{90}{\emph{Average}} \\
    \midrule
    
    Co-CoOp & 99.99 & 99.98 & 99.99 & 100.00 & 99.97 & 99.96 & 99.98 & 100.00 & 99.36 & 99.80 & 100.00 & 99.91 \\
    MaPLe & 99.99 & 99.92 & 100.00 & 100.00 & 100.00 & 99.94 & 100.00 & 99.97 & 98.76 & 100.00 & 99.84 & 99.86 \\
    
 PromptSRC & 99.98 & 100.00 & 100.00 & 99.98 & 100.00 & 99.96 & 100.00 & 99.89 & 99.88 & 100.00 & 99.71 & 99.95 \\
    \bottomrule
    \end{tabular}
    }
    \vspace{-1em}
    \label{tab:xd_WSR} 
\end{table*}

\vspace{0.3em}
\noindent\textbf{Settings for Prompt Watermarking.} We set margin hyperparameter $\varepsilon$ to 0.5 and the loss hyperparameter $\lambda$ to 1. \blue{For watermark embedding, we use four verification classes named ``Target 1", ``Target 2", ``Target 3", and ``Target 4". Note that this naming choice is a default setting. In Section~\ref{name}, we empirically demonstrate that the performance of our method remains robust even when the verification class names have semantic overlap with the original task classes.} For our BWAP methods, the verification class is ``Target" and all watermarking settings are set in the same way as its original settings \citep{gu2017badnets,nguyen2021wanet}.

\begin{table*}[!t]
    \centering
    \caption{\textnormal{Accuracy on domain generalization. }} 
        % \vspace{-0.8em}
 \setlength{\tabcolsep}{8pt}
    % \scalebox{0.75}[0.75]{
    \begin{tabular}{l cccccc}
    \toprule
    & \textbf{Source} & \multicolumn{5}{c}{\textbf{Target}} \\ \cmidrule(lr){2-2} \cmidrule(lr){3-7}
     & ImageNet & -V2 & -S & -A & -R  & Avg.\\
    \midrule
    Co-CoOp & 70.90 & 63.71 & 48.18 & 49.91 & 76.28 & 61.80  \\
    MaPLe & 70.55 & 63.48 & 48.71 & 50.59 & 75.95 & 61.86  \\
    PromptSRC & 70.53 & 63.90 & 49.23 & 50.87 & 77.49 & 62.40 \\
    \bottomrule
    \end{tabular}
    % }
    \label{tab:domain_acc}
    \vspace{-2em}
\end{table*}

\begin{table*}[!t]
    \centering
    \caption{\textnormal{WSR on domain generalization. }}
        % \vspace{-0.8em}
 \setlength{\tabcolsep}{8pt}
    % \scalebox{0.75}[0.75]{
    \begin{tabular}{l cccccc}
    \toprule
    & \textbf{Source} & \multicolumn{5}{c}{\textbf{Target}} \\ \cmidrule(lr){2-2} \cmidrule(lr){3-7}
     & ImageNet & -V2 & -S & -A & -R  & Avg.\\
    \midrule
    Co-CoOp & 99.99 & 99.90 & 99.83 & 99.89 & 99.79 & 99.88  \\
    MaPLe & 99.99 & 99.92 & 99.85 & 99.97 & 99.92 & 99.93  \\
    PromptSRC & 99.98 & 99.89 & 99.93 & 99.77 & 99.89 & 99.89 \\
    \bottomrule
    \end{tabular}
    % }
    % \vspace{-1em}
         
    \label{tab:domain_wsr}
    
\end{table*}

\vspace{0.3em}
\noindent\textbf{Settings for Ownership Verification.} \blue{We randomly select $m=100$ different test samples from novel classes for hypothesis testing. Note that $m=100$ is a relatively conservative choice. In Section~\ref{queries}, we empirically demonstrate that our method remains effective with significantly fewer test samples, further confirming its practicality in real world scenarios.} Each test repeats three times using all selected samples and we calculate the average p-value to reduce the impact of randomness. The verification classes are the same as the verification classes embedded in prompt watermarking. The significance level $\alpha$ is set to 0.01 and the threshold parameter $\tau$ is set to 0.5. Detailed experimental settings are presented in Appendix \ref{set_main}.
%\blue{We include all original test classes together with the verification classes during ownership verification process. The primary purpose is to verify that embedding the watermark does not degrade downstream task performance, thereby validating the harmlessness requirement of our method. In real-world scenarios, the verifier can simply input a small number of candidate classes (\eg, 3 verification classes and 2 task-relevant classes), since commercial APIs may only return top-5 predictions. Such a setting is sufficient for the verifier to both confirm correct classification of task-relevant classes and check the sequential ordering of verification classes for ownership verification.} 

\begin{table*}[t!]
    \centering
    \caption{Watermarking and verification performance of SWAP. We hereby report results for three settings: the watermarked model (\ie, `Watermarked'), an independently trained prompt (\ie, `Ind Prompt'), and independent verification classes (\ie, `Ind Veri').}
    % \vspace{-0.8em}
    % \resizebox{0.47\textwidth}{!}{
     \setlength{\tabcolsep}{3pt}
        \begin{tabular}{c|c|cc |cc |cc}
        \toprule
          & Dataset{$\rightarrow$}
         & \multicolumn{2}{c}{ImageNet}
         & \multicolumn{2}{c}{Caltech101} &
          \multicolumn{2}{c}{OxfordPets} 
         % &\multicolumn{5}{c}{Food101}
         \\
        
           \cmidrule(lr){3-4}  \cmidrule(lr){5-6} \cmidrule(lr){7-8} 
           %\cmidrule(lr){18-22}

        \makecell{Prompt Tuning \\Method}& \makecell{Scenario$\downarrow$}
        &  WSR     &p-value  & WSR     &p-value & WSR     &p-value 
        % & Base    & Novel  & WSR     &p-value &$\hat{H}$
        \\
       \midrule
        \multirow{3}{*}{\makecell{CoCoOp}} 
        & Ind Prompt & 4.27 & 1 & 2.62 &  1 & 7.10 & 1 
        % & 90.70& 91.29& 1.92&  0.70 & -
        \\ 
        & Ind Veri & 1.30 & 1 & 1.20 &  1 & 3.52 & 1 
        % & 0.70& 99.60 & 0.70&  0.70 & -
        \\ 
        & Watermarked & 99.98 & 0 & 99.67 &  0 & 99.36 & 0 
        %& 88.90& 89.80 & 98.40& 99.60 & 0.90
        \\ 
        \midrule
        \multirow{3}{*}{MaPLe} 
        & Ind Prompt & 3.20 &1& 1.53 & 1 & 3.30 & 1
        %& 90.71& 92.05& 3.61 &  0.70 & -
        \\ 
        & Ind Veri & 2.86 & 1 & 4.04 &  1 & 3.69 & 1 
        %& 0.70& 99.60 & 0.70&  0.70 & -
        \\ 
        & Watermarked & 99.95 & 0 & 99.99 &  0 & 99.94 & 0
        %& 89.80& 90.60& 99.70 & 0.70& 0.92
        
        \\ 
        \midrule
        \multirow{3}{*}{PromptSRC} 
        & Ind Prompt & 3.01& 1&5.90 &  1 & 1.40 & 1 
        %& 90.67& 91.53& 4.10&  0.70 & -
        \\ 
        & Ind Veri & 2.38 & 1 & 4.69 &  1 & 6.82 & 1 
        %& 0.70& 99.60 & 0.70&  0.70 & -
        \\ 
        & Watermarked & 99.97 & 0 & 99.94 &  0 & 99.99 & 0 
        %& 90.30& 91.10& 99.60 & 0.70& 0.91
        
        \\ 
        % \midrule
        % \multirow{3}{*}{ATPrompt} 
        % & Ind Prompt & 3.41& 1&2.15 &  1 & 7.69 & 1 
        % %& 90.67& 91.53& 4.10&  0.70 & -
        % \\ 
        % & Ind Veri & 4.92 & 1 & 3.06 &  1 & 1.12 & 1 
        % %& 0.70& 99.60 & 0.70&  0.70 & -
        % \\ 
        % & Watermarked & 100.00 & 0 & 100.00 &  0 & 100.00 & 0 
        % %& 90.30& 91.10& 99.60 & 0.70& 0.91
        
        % \\ 

       \bottomrule
    \end{tabular}
    % }
    % \vspace{-1.5em}
    \label{ind_res}
\end{table*}

\subsection{Results under Benign Users}
\noindent \textbf{Effectiveness of SWAP in Base-to-novel Scenario.} As shown in Table \ref{main_res}, our SWAP achieves the best performance in all cases. For effectiveness, the p-values of our method are far less than the significance level $\alpha$ and the WSRs are nearly equal to 1, demonstrating the effectiveness of our prompt watermarking and verification, whereas all other CLIP-based backdoor methods have very low WSRs, proving they cannot be effectively adapted to this scenario as mentioned in Section \ref{limi_back}. For harmlessness, our method maintains the same high accuracy on both base and novel classes as the original prompt tuning method, while the BWAP shows slight decline in both metrics. Additionally, the $\hat{H}$ of our method approaches 0, whereas nearly all BWAP's values exceed 0.7, indicating our accuracy remains unaffected during verification. These results demonstrate the harmlessness of our approach. Detailed results across eleven datasets, and results using the SOTA prompt tuning method ATPrompt \citep{li2024advancing}, are provided in Appendix \ref{add_exp}.

\vspace{0.3em}
\noindent \textbf{Effectiveness of SWAP in Cross-dataset Scenario. } Having demonstrated the effectiveness of SWAP within a single dataset, we extend our evaluation to the cross-dataset scenario. In this setting, the model is trained entirely on a single source dataset and evaluated on entirely different target datasets. This represents a more common and critical real-world application, where prompt developers train on  private datasets and publicly release the soft prompt for prompt reusers to apply across diverse target datasets. In cross-dataset transfer scenario, we train the prompt on ImageNet and directly evaluate it on other datasets without any data-specific fine-tuning. As shown in Table \ref{tab:xd_acc}, our method demonstrates high accuracy across various prompt tuning methods on both source and target datasets, confirming the harmlessness of our approach in cross-dataset scenarios. Furthermore, as shown in Table \ref{tab:xd_WSR}, our method exhibits high WSRs on both source and target datasets, validating the effectiveness of prompt watermarking in cross-dataset scenarios. 
%进一步讨论，通过可以保证在一个数据集训练，在别的数据集
%进一步讨论，不同数据集，但是domain是相同等domain distribution shifting的场景

\begin{table*}[t!]
    \centering
    \caption{The resistance to false claim attacks of our method with CLIP as the reference model and various prompt tuning methods as victim models.}
    % \vspace{-0.8em}
    % \resizebox{0.47\textwidth}{!}{
     \setlength{\tabcolsep}{3pt}
        \begin{tabular}{c|c|cc|cc}
        \toprule
        Dataset& \makecell{REF Model \& \\VICTIM Model}& REF ACC.   & REF ASR  & VICTIM ACC.  & VICTIM ASR  \\
       \midrule
        \multirow{3}{*}{\makecell{ImageNet}} 
        & CLIP$\rightarrow$CoOp & \multirow{3}{*}{68.14} & \multirow{3}{*}{4.16} & 67.88 & 9.76\\ 
        & CLIP$\rightarrow$CoCoOp &&& 70.43 & 11.73\\ 
        & CLIP$\rightarrow$MaPLe &&& 70.54 & 13.42\\ 
        \midrule
        \multirow{3}{*}{Caltech101} 
        & CLIP$\rightarrow$CoOp & \multirow{3}{*}{94.00} & \multirow{3}{*}{13.11} & 89.81 & 4.38\\ 
        & CLIP$\rightarrow$CoCoOp &&& 93.81 & 10.81\\ 
        & CLIP$\rightarrow$MaPLe &&& 94.36 & 7.21\\ 
        \midrule
        \multirow{3}{*}{OxfordPets} 
        & CLIP$\rightarrow$CoOp & \multirow{3}{*}{97.26} & \multirow{3}{*}{11.92} & 95.29 & 8.63\\ 
        & CLIP$\rightarrow$CoCoOp &&& 97.69 & 9.74\\ 
        & CLIP$\rightarrow$MaPLe &&& 97.76 & 13.30\\ 
        \midrule
        \multirow{3}{*}{Food101} 
        & CLIP$\rightarrow$CoOp & \multirow{3}{*}{91.22} & \multirow{3}{*}{7.30} & 82.26 & 6.58\\ 
        & CLIP$\rightarrow$CoCoOp &&& 91.29 & 16.37\\ 
        & CLIP$\rightarrow$MaPLe &&& 92.05 & 11.46\\ 
       \bottomrule
    \end{tabular}
    % }
    \vspace{-1em}
    \label{resis_false}
\end{table*}

\vspace{0.3em}
\noindent \textbf{Effectiveness of SWAP in Domain Generalization Scenario. } Furthermore, we conduct experiments on domain generalization scenarios, where the data distributions between the source and target datasets exhibit a significant shift, to prove SWAP’s broader effectiveness. In domain generalization scenario, we train the prompt on ImageNet and evaluate on out-of-distribution datasets, including ImageNet-A \citep{hendrycks2021natural}, ImageNet-R \citep{hendrycks2021many}, ImageNet-Sketch \citep{wang2019learning} and ImageNetV2 \citep{recht2019imagenet}, 
to test performance under domain shifts. As illustrated in Table \ref{tab:domain_acc}, our approach achieves excellent accuracy with different prompt tuning methods for both source and target datasets, verifying that our method remains benign in domain generalization scenarios. Additionally, Table \ref{tab:domain_wsr} reveals that our technique maintains robust WSRs across source and target datasets, establishing the effectiveness of prompt watermarking under domain generalization scenarios. 

\vspace{0.3em}
\noindent \textbf{Distinctiveness of SWAP. } Table \ref{ind_res} demonstrates the distinctiveness of SWAP compared to independently trained prompts and independently verification classes. The WSRs in both scenarios are all below 10\% and the p-values are large, indicating that prompts without embedding of the watermark cannot exhibit the features of the watermark. Furthermore, WSRs in both scenarios are approximately 4\%, which closely approximates the probability of one specific permutation among all possible arrangements of 4 verification classes ($\frac{1}{4!} =\frac{1}{24}\approx 4.17\%$). This indicates that without watermark embedding, the arrangement of these classes is completely random. 
% The more ablation studies of three key hyperparameters (\ie, $\mathit{n}$, $\lambda$, and $\varepsilon$) and more detailed results across eleven image recognition datasets are provided in Appendix \ref{add_exp}.

\begin{table*}[t!]
    \centering
    \vspace{-1em}
    \caption{The results of resistance to adaptive false claim attacks of our method, where A, B, and C represent CoCoOp, MaPLe, and PromptSRC, respectively.}
% \vspace{-0.8em}
    \resizebox{0.95\textwidth}{!}{
     \setlength{\tabcolsep}{3pt}
        \begin{tabular}{c|cccc|cc}
        \toprule
        Attack Settings& REF1 ACC.  & REF1 ASR  & REF2 ACC.  & REF2 ASR  & VICTIM ACC.   
         & VICTIM ASR    \\
       \midrule
       A $\&$ B$\rightarrow$C& 93.81& 88.88& 94.36& 98.38& 94.03& 7.21\\
       \midrule
       B $\&$ C$\rightarrow$A& 94.36& 96.75& 94.03& 94.12& 93.81& 4.59\\
       \midrule
       A $\&$ C$\rightarrow$B& 93.81& 91.88& 94.03& 89.25& 94.36& 6.22\\
       
        \bottomrule
    \end{tabular}
    }
    
    \vspace{-1em}
    \label{ada_resis_false}
\end{table*}

\begin{figure*}[t!]
    \centering
    \begin{minipage}[t]{0.24\linewidth}
        \centering
        \includegraphics[width=1.00\linewidth]{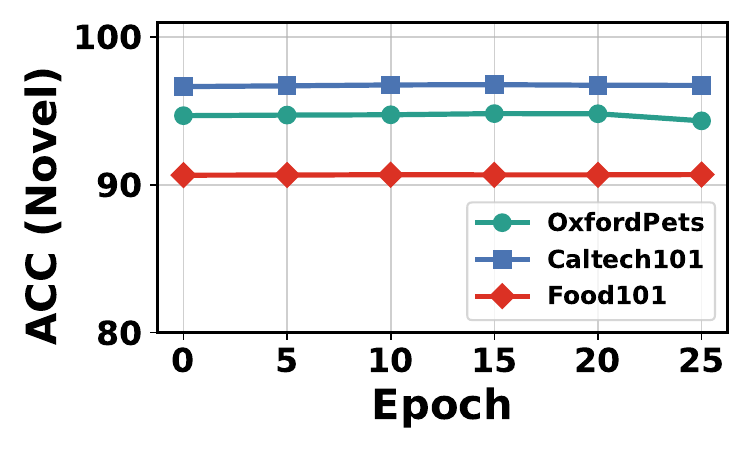}
    \end{minipage}
    \begin{minipage}[t]{0.24\linewidth}
        \centering
        \includegraphics[width=1.00\linewidth]{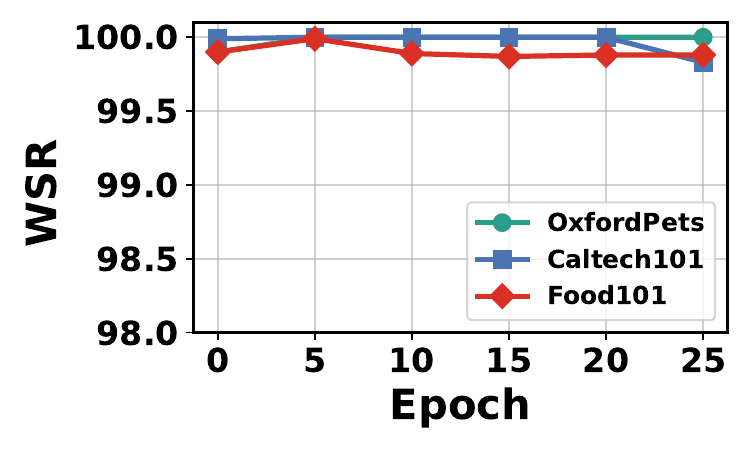}
    \end{minipage}
    \begin{minipage}[t]{0.24\linewidth}
        \centering
        \includegraphics[width=1.00\linewidth]{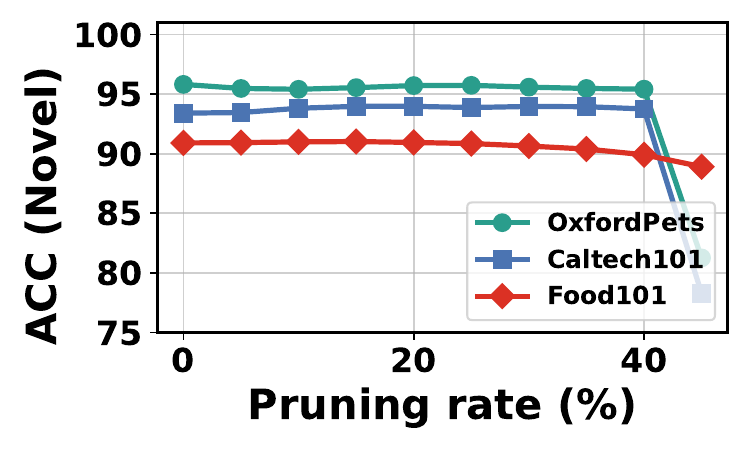}
    \end{minipage}
    \begin{minipage}[t]{0.24\linewidth}
        \centering
        \includegraphics[width=1.00\linewidth]{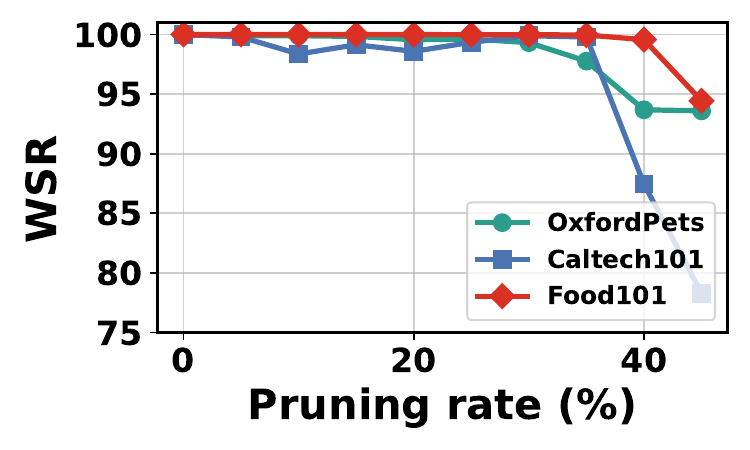}
    \end{minipage}
    \caption{The resistance to fine-tuning Attack (left) and model-pruning attack (right).}
    \vspace{-1em}
    \label{fig:attacks}
\end{figure*}

\subsection{Results under False Claim Attacks}

We hereby follow the same settings as the false claim attack against BWAP in Section \ref{limitation_bwap}, except that the adversarial objective is the same loss as E.q. (\ref{order}). As shown in Table \ref{resis_false}, The adversarial examples achieve a low attack success rate (ASR) on the reference model, not to mention the victim model. This demonstrates SWAP's strong resistance to false claim attacks. We argue that adversarial examples achieve low success rates on the victim model because CLIP is used as the reference model and has no tunable prompts. Consequently, the attack only involves a simple, step-wise update of the adversarial perturbation in the gradient direction. However, our watermarking objective involves only minimal, sequential adjustments for out-of-distribution verification classes in the embedding space, which is too subtle for the step-wise gradient updates used by adversarial perturbation to effectively target. More challenging settings with tunable prompts in the reference model and multiple reference models for false claim attacks will be discussed in Section \ref{ada_att}.

% As shown in Table \ref{resis_false}, SWAP demonstrates strong resistance to false claim attacks. The adversarial examples achieve low attack success rates on both the reference model and victim model, indicating the difficulty in generating adversarial examples that satisfy these objectives. Since CLIP is used as the reference model and it has no tunable prompts, adversarial attacks cannot achieve good results on the reference model either. We argue that adversarial examples achieve low attack success rates on both the reference and victim models because our watermarking objective, which involves only minimal, sequential adjustments for out-of-distribution verification classes, is too subtle for the step-wise gradient updates used by adversarial examples to effectively target. More challenging settings with tunable prompts in the reference model and multiple reference models for false claim attacks will be discussed in Section \ref{ada_att}.

%上 as shwon in 9，reference很差，更不用说victim。证明了xxx。说明victim的原因是

\subsection{Resistance to Potential Attacks} 
We hereby evaluate the resistance of our SWAP to potential attacks. Specifically, we consider five types of attacks: finetuning attacks, model pruning attacks, adaptive false claim attacks, overwriting attacks, and unlearning attacks. 

\vspace{0.3em}
\noindent\textbf{The Resistance to Fine-tuning Attack.} We evaluate the robustness of our method against fine-tuning attacks \citep{liu2017neural} by training the watermarked model on a local benign dataset for several epochs. The prompt tuning method is MaPLe. Figure \ref{fig:attacks} demonstrates that our method maintains stable WSR and ACC (Novel) under fine-tuning attacks. We argue that this robustness primarily results from preserving original labels during watermarking, unlike backdoor-based approaches that alter sample labels.

%统一名字ACC (Base) ACC (Novel)

\vspace{0.3em}
\noindent \textbf{The Resistance to Model-pruning Attack.} Model pruning \citep{han2015learning} challenges watermark robustness through the potential elimination of watermark-related neurons. We conduct the experiments on MaPLe. As shown in Figure \ref{fig:attacks}, the ACC (Novel) decreases with the decrease of WSR, demonstrating that our method is resistant to model pruning attack.
\label{ada_att}

\vspace{0.3em}
\noindent \textbf{The Resistance to Adaptive False Claim Attack.} We employ two prompt tuning methods as reference models while using another prompt tuning method as the victim model. The adversarial objectives are formulated as:
\begin{equation}
\begin{gathered}
\min_{\theta_1, \theta_2} \mathcal{L}_a = \mathcal{L}'_{f} + \lambda \mathcal{L}'_{o}, 
    \\
    \bar{\bm{x}}_{t+1} =\mathit{Clip}_{\bm{x},\epsilon}(\bar{\bm{x}}_t - \gamma \cdot sign(\nabla \mathcal{L}_a(\theta_1, \theta_2 ;y,\bar{\bm{x}}_t)  )),
\end{gathered}
\end{equation}
where $\mathit{Clip}_{\bm{x},\epsilon}(\cdot)$ constrains the perturbation within $\epsilon$ under the $L_\infty$ norm, $\theta_1$ and $\theta_2$ are the parameters of two reference prompts, 
$\gamma$ is the step size and $\nabla \mathcal{L}_a(\theta_1, \theta_2 ;y,x'_t)$ computes the gradient of the loss function $\mathcal{L}_a$ on two prompts. We conduct the experiments on Caltech101. As shown in Table \ref{ada_resis_false}, SWAP demonstrates strong resistance to adaptive false claim attacks. While the adversarial examples achieve high attack success rates on both reference models, they fail to transfer successfully to the independent victim model, indicating minor transferability of adversarial examples under this more challenging scenario.

\vspace{0.3em}
\noindent \textbf{The Resistance to Overwriting Attack.} In this attack scenario, we consider an adversary who is familiar with the SWAP methodology but lacks knowledge about the specific verification classes chosen by the original prompt developer. Consequently, the attacker may attempt to inject their own set of verification classes into the prompt, with the goal of overwriting or invalidating the previously embedded watermark. To assess resistance to overwriting attacks, we performed fine-tuning on the prompt for 5 epochs using the original objective but with 4 completely different verification classes. Specifically, the verification classes embedded by the prompt developer are ``Target 1", ``Target 2", ``Target 3", and ``Target 4", while the adversary attempted to embed ``Miqi 1", ``Miqi 2", ``Miqi 3", and ``Miqi 4". The results of this overwriting attack are shown in Table \ref{more_ada_att}. We found that even after embedding new verification classes, the WSR of the original classes remained at 100\%, demonstrating that the original watermark could not be overwritten. Interestingly, we also observed that the WSR of the newly embedded verification classes was quite high. However, this is a common and trivial situation in watermarking. This issue can be solved by registering the watermark and the prompt to a trusted third party accompanied by timestamps. The watermark with a later timestamp will not be treated as a valid copyright certificate \citep{waheed2024grove}. These findings confirm that our method effectively resists overwriting attacks.

\begin{table*}[t!]
    \centering
    \caption{The resistance to overwriting and unlearning attacks.}
    % \vspace{-0.8em}
    % \resizebox{0.47\textwidth}{!}{
     \setlength{\tabcolsep}{3pt}
        \begin{tabular}{c|cccc}
        \toprule
        \makecell{Prompt Tuning \\Method }& Metric & Before & After Overwriting& After Unlearning\\
       \midrule
       \multirow{5}{*}{\makecell{CoCoOp}}&ACC (base) (\%)&97.65&97.74&97.42\\
       &ACC (Novel) (\%)&93.10&93.01&92.9\\
       &p-value&0&0&0\\
       &WSR with Ori WM (\%)&99.67&99.68&99.94\\
       &WSR with New WM (\%)&-&96.97&-\\
       \midrule
      \multirow{5}{*}{\makecell{MaPLe}}&ACC (base) (\%)&97.30&97.42&97.13\\
       &ACC (Novel) (\%)&95.31&94.54&94.49\\
       &p-value&0&0&0\\
       &WSR with Ori WM (\%)&99.99&100.00&100.00\\
       &WSR with New WM (\%)&-&98.77&-\\
       \midrule
       \multirow{5}{*}{\makecell{PromptSRC}}&ACC (base) (\%)&96.28&96.03&95.71\\
       &ACC (Novel) (\%)&94.21&93.89&94.32\\
       &p-value&0&10$^{-17}$&0\\
       &WSR with Ori WM (\%)&99.94&99.23&99.94\\
       &WSR with New WM (\%)&-&97.55&-\\
        \bottomrule
    \end{tabular}
    % }
    
    %\vspace{-1em}
    \label{more_ada_att}
\end{table*}
\begin{table*}[!t]
\centering
\captionsetup{labelfont={color=tablecolor}, textfont={color=tablecolor}}
\caption{Watermarking and verification performance of MaPLe under a Gaussian logit perturbation attack. 
Standard normal noise $\mathcal{N}(0,1)$ (std = 1.0, without additional scaling) is injected into all non-top-1 logits, while the top-1 prediction is preserved via masking.}
% \vspace{-0.3em}
\resizebox{\linewidth}{!}{
\begin{tabular}{>{\color{tablecolor}}c >{\color{tablecolor}}c>{\color{tablecolor}}c>{\color{tablecolor}}c >{\color{tablecolor}}c>{\color{tablecolor}}c>{\color{tablecolor}}c >{\color{tablecolor}}c>{\color{tablecolor}}c>{\color{tablecolor}}c}
\toprule
Dataset$\rightarrow$ 
& \multicolumn{3}{c}{\blue{ImageNet}} 
& \multicolumn{3}{c}{\blue{Caltech101}} 
& \multicolumn{3}{c}{\blue{OxfordPets}} \\
\cmidrule(lr){2-4} \cmidrule(lr){5-7} \cmidrule(lr){8-10}

Protection Method 
& ACC & WSR & p-value 
& ACC & WSR & p-value 
& ACC & WSR & p-value \\
\midrule

{SWAP (w/ Gaussian Noise)}
& 77.06& 86.19 & $3 \times 10^{-57}$
& 97.62 & 89.31 & $5 \times 10^{-17}$
& 94.77 & 85.96 & $3\times 10^{-14}$ \\

\bottomrule
\end{tabular}
}
\label{gau}
\vspace{-1em}
\end{table*}

\vspace{0.3em}
\noindent \textbf{The Resistance to Unlearning Attack.} In this attack scenario, we examine a more sophisticated adversary who possesses knowledge of both the SWAP methodology and the specific verification classes embedded within the prompt. With this information, the attacker attempts to deliberately remove the watermark by performing gradient updates in the opposite direction of the watermarking objective. This approach, known as an unlearning attack, aims to systematically degrade the watermark's effectiveness while preserving the prompt's original functionality. The adversary uses the following loss function to unlearn the watermark:
\begin{equation}
\min_{\theta_{V_f},\theta_{V_g}} \mathcal{L}_u = \mathcal{L''}_{f} - \lambda \mathcal{L''}_{o}.
\label{unlearn_loss}
\end{equation}
We performed fine-tuning on the prompt for 5 epochs using unlearning loss. Specifically, the verification classes embedded by the prompt developer are ``Target 1", ``Target 2", ``Target 3", and ``Target 4". The experimental results of the unlearning attack are shown in Table \ref{more_ada_att}. The findings show that our SWAP method effectively resists unlearning attacks. Specifically, the WSRs show a negligible decrease from their original values. The watermark can still be successfully extracted from the model, and ownership can still be verified with statistically significant low p-values.

{\color{blue}
\begin{table*}[!t]
\centering
\captionsetup{labelfont={color=tablecolor}, textfont={color=tablecolor}}
\caption{Watermarking and verification performance of MaPLe under semantically overlapping verification settings on different datasets. 
(a) Verification classes selected from testing-set categories that are semantically close to other task classes. 
(b) Verification classes constructed as fine-grained variants of existing task categories.}
\vspace{-0.3em}
\resizebox{\linewidth}{!}{
\begin{tabular}{>{\color{tablecolor}}l >{\color{tablecolor}}c>{\color{tablecolor}}c>{\color{tablecolor}}c >{\color{tablecolor}}c>{\color{tablecolor}}c>{\color{tablecolor}}c >{\color{tablecolor}}c>{\color{tablecolor}}c>{\color{tablecolor}}c}
\toprule
Dataset$\rightarrow$ & \multicolumn{3}{c}{\blue{ImageNet}} & \multicolumn{3}{c}{\blue{Caltech101}} & \multicolumn{3}{c}{\blue{OxfordPets}} \\
\cmidrule(lr){2-4} \cmidrule(lr){5-7} \cmidrule(lr){8-10}
Protection Method & ACC & WSR & p-value & ACC & WSR & p-value & ACC & WSR & p-value \\
\midrule

SWAP (Testing-overlap) 
& 76.97 & 99.93 & 0
& 97.48 & 100 & 0 
& 94.79 & 100 & 0 \\

SWAP (Sub-category) 
& 77.05 & 99.96 & 0
& 96.45 & 99.94 & 0 
& 94.96 & 100 & 0 \\

\bottomrule
\end{tabular}
}
\label{seman}
\end{table*}
}

\begin{figure*}[t!]
    \centering
    \begin{minipage}[t]{0.24\linewidth}
        \centering
        \includegraphics[width=1.00\linewidth]{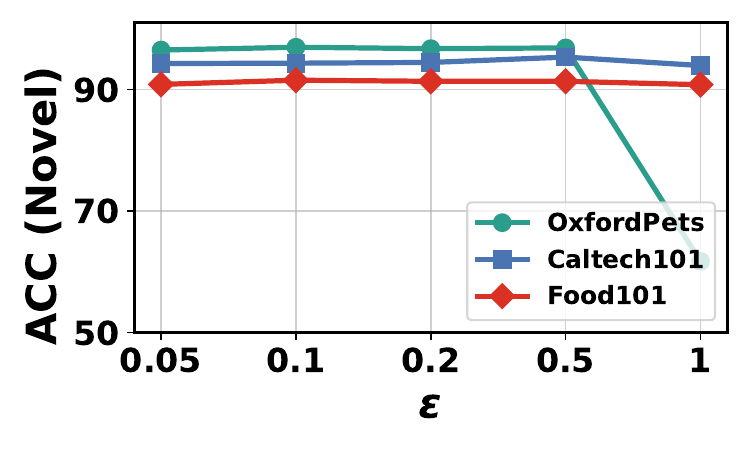}
    \end{minipage}
    \begin{minipage}[t]{0.24\linewidth}
        \centering
        \includegraphics[width=1.00\linewidth]{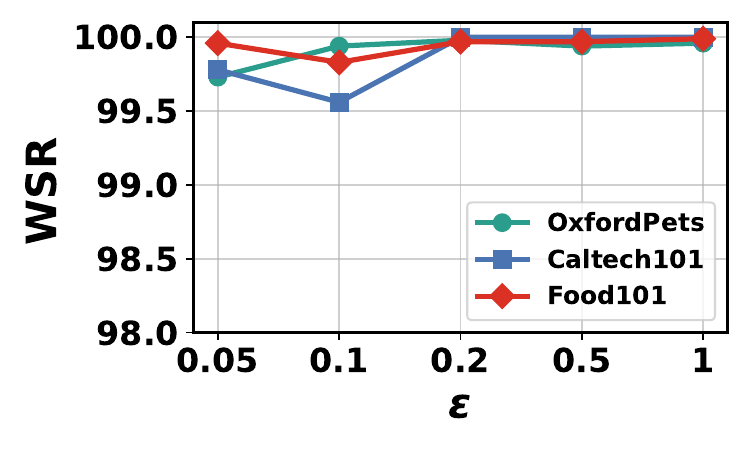}
    \end{minipage}
    \begin{minipage}[t]{0.24\linewidth}
        \centering
        \includegraphics[width=1.00\linewidth]{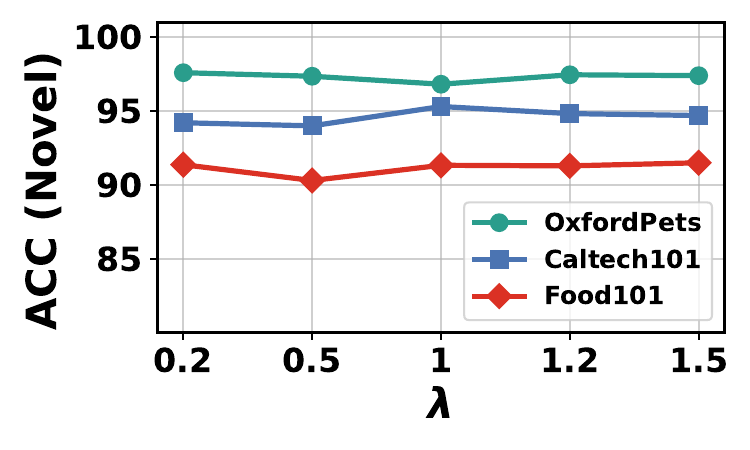}
    \end{minipage}
    \begin{minipage}[t]{0.24\linewidth}
        \centering
        \includegraphics[width=1.00\linewidth]{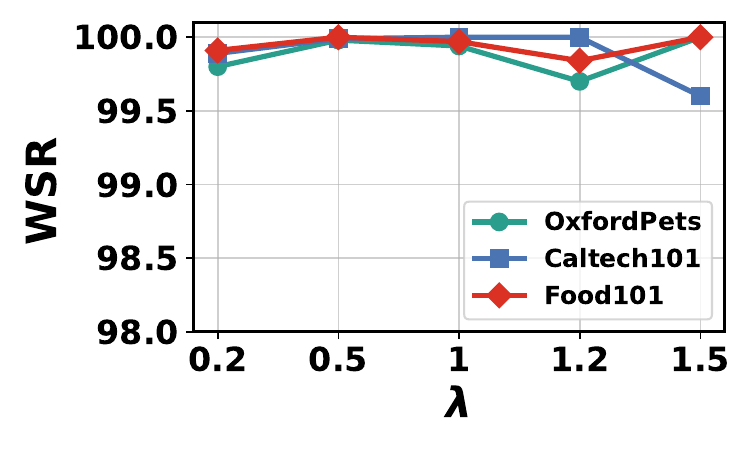}
    \end{minipage}
    \caption{The sensitivity of hyper-parameter $\varepsilon$ (left) and $\lambda$ (right).}
    \vspace{-1em}
    \label{fig:para}
\end{figure*}

\vspace{0.3em}
\noindent \blue{\textbf{The Resistance to Low-logit Perturbation Attack.} In this attack scenario, we consider an adversary who is familiar with the SWAP methodology but lacks knowledge of the specific verification classes selected by the original prompt developer. Since the attacker cannot identify which classes serve as verification classes, a realistic strategy is to add Gaussian noise to the logits of all non-top-1 categories in an attempt to disrupt the predefined probability ranking used for ownership verification. Specifically, at inference time, we add Gaussian noise with standard deviation $\sigma=1.0$ to the output logits of all categories except the top-1 predicted class, ensuring that the model's primary prediction remains unaffected while potentially disturbing the ordering among lower-ranked classes. As shown in Table~\ref{gau}, both ACC and WSR maintain strong performance under perturbation, demonstrating that SWAP is robust against low-logit perturbation attack.}

\subsection{Ablation Studies}
We hereby discuss the effects of three key hyperparameters involved in our method (\ie, $\mathit{n}$, $\lambda$, and $\varepsilon$). We study their effects on Caltech101, OxfordPets, and Food101 datasets with MaPLe as the prompt tuning method.

\begin{figure}[t!]
    \centering
    \begin{minipage}[t]{0.48\linewidth}
        \centering
        \includegraphics[width=1\linewidth]{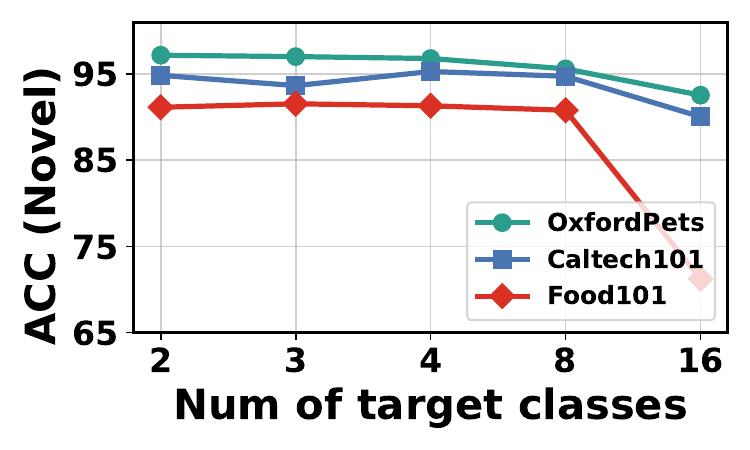}
    \end{minipage}
    \begin{minipage}[t]{0.48\linewidth}
        \centering
        \includegraphics[width=1\linewidth]{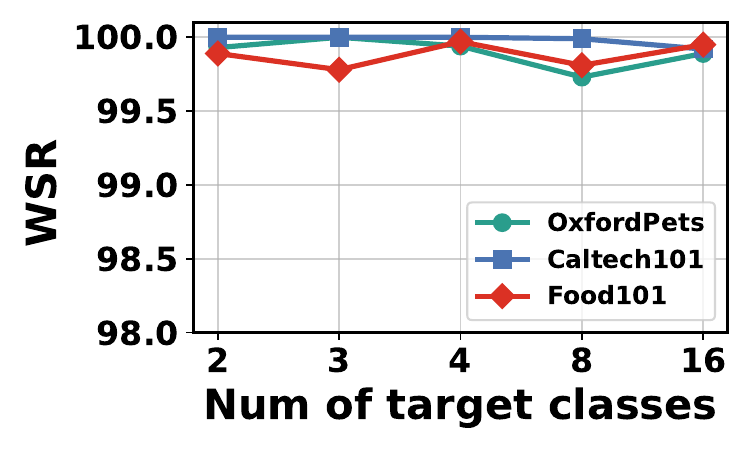}
    \end{minipage}
    % \vspace{-3mm}
    \caption{Effect of the Num of Target Classes.}

    \label{fig:numclass}
    \vspace{-1em}
 \end{figure}

\vspace{0.3em}
\noindent \textbf{Effect of the Num of Target Classes.} As shown in Figure \ref{fig:numclass}, ACC (Novel) and WSR maintain excellent performance as the number of target classes increases, demonstrating that our method can ensure stronger verification capabilities through an expanded set of target classes.

\vspace{0.3em}
\noindent \textbf{Effect of the Hyperparameter $\lambda$.} As shown in Figure \ref{fig:para}, ACC (Novel) and WSR maintain its performance as $\lambda$ increases, demonstrating the robustness of our method.

\vspace{0.3em}
\noindent \textbf{Effect of the Margin Hyperparameter $\varepsilon$.} As shown in Figure \ref{fig:para}, WSR maintains excellent performance as $\varepsilon$ increases, while ACC experiences only a slight decline. Notably, both ACC (Novel) and WSR achieve strong performance even with small $\varepsilon$ (e.g., 0.05), indicating that our method can utilize more target classes to ensure stronger verification capabilities.

\begin{figure}[!t]
    \centering
    \includegraphics[width=0.8\linewidth]{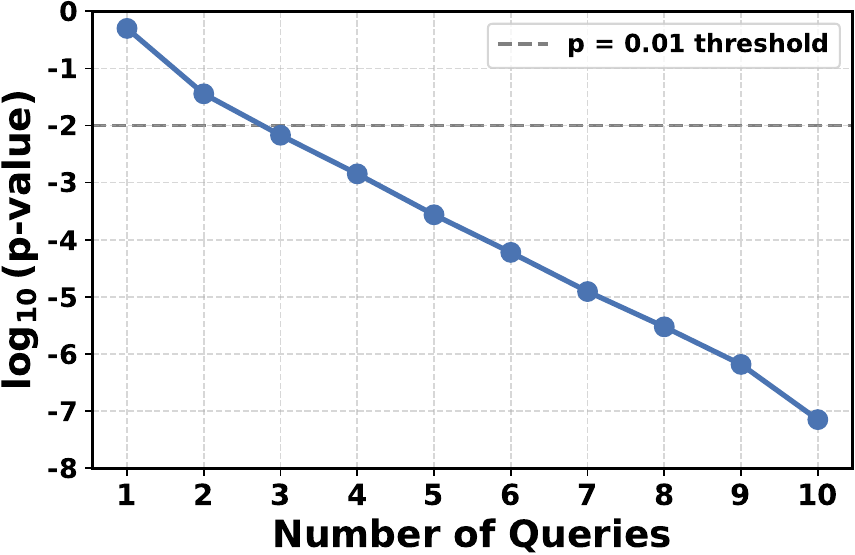}
    \vspace{-0.3em}
    \captionsetup{labelfont={color=tablecolor}, textfont={color=tablecolor}}
    \caption{Effect of the number of verification queries on the p-value. }
    \label{queries-pvalue}
    \vspace{-1em}
\end{figure}

\vspace{0.3em}
\label{name}
\noindent \blue{\textbf{Effect of the Selection of Verfication Classes. }To evaluate the robustness of SWAP under semantically overlapping verification classes, we consider two challenging scenarios: (a) \underline{Testing-set classes as verification classes}: We select a subset of classes from the training set as verification classes, which are semantically close to other training/testing classes. Specifically, the verification classes are \textit{"nautilus", "trilobite", and "llama"} for Caltech101, \textit{"newfoundland", "persian", and "pomeranian"} for OxfordPets, and \textit{"cloak", "clogs", and "spindle"} for ImageNet. (b) \underline{Fine-grained sub-categories as verification classes}: We use sub-categories of existing task classes as verification classes. Specifically, the verification classes are \textit{"camera"$\rightarrow$"brokencamera", brain"$\rightarrow$"smartbrain", and "face"$\rightarrow$"smilingface"} for Caltech101, \textit{"Beagle"$\rightarrow$"HappyBeagle", "Bengal"$\rightarrow$"CoolBengal", and "Havanese" $\rightarrow$ "SweetHavanese"} for OxfordPets, and \textit{"Jay"$\rightarrow$"NoisyJay", "Magpie"$\rightarrow$"SmartMagpie", and "Chickadee"$\rightarrow$"Cute
Chickadee"} for ImageNet. As shown in Table~\ref{seman}, both ACC and WSR maintain strong performance across the two scenarios, demonstrating that our method preserves both high effectiveness and stealthiness even when verification categories have high semantic overlap with other candidate classes.}

\vspace{0.3em}
\noindent \blue{\textbf{Effect of the Number of Verfication Queries. }As shown in Figure~\ref{queries-pvalue}, the p-value decreases as the number of verification queries increases. With as few as 5 queries, the p-value already falls below the significance level $\alpha=0.01$. This result demonstrates that the choice of $m=100$ in our experimental settings is a relatively conservative one, primarily intended to provide stronger statistical evidence for demonstrating the effectiveness of our method. In practice, our method can successfully perform verification with as few as 10 queries, which incurs negligible cost to the verification process.}
\label{queries}

\begin{figure}[!t]
 %\vspace{-1em}
  \centering
\includegraphics[width=1\linewidth]{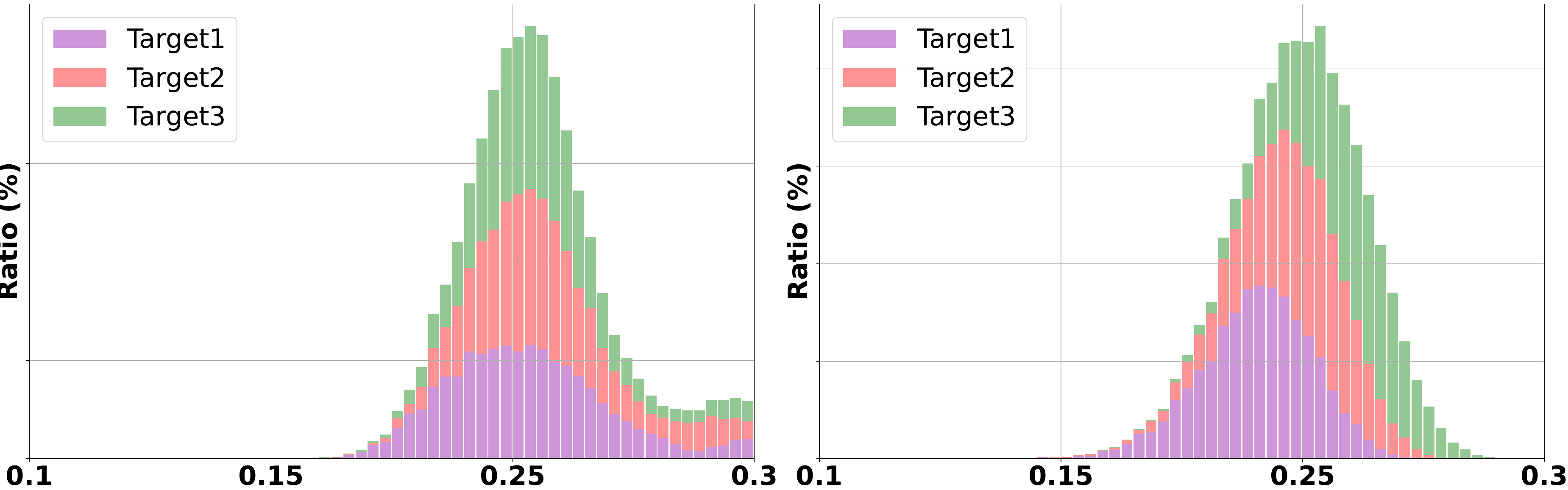}
\vspace{0.2em}
  \caption{Distribution of cosine similarities between images and three verification classes in the feature space. While the independent model $\mathbf{(left)}$ exhibits similar distributions of cosine similarities across the three verification classes, the watermarked model $\mathbf{(right)}$ demonstrates a consistent rightward shift in these distributions, suggesting a progressive increase in cosine similarities from ``Target 1'' to ``Target 
 3''.}
  \vspace{-1em}
  \label{fig:dis}
\end{figure}

\subsection{Towards A Deeper Understanding of SWAP}
\begin{figure}[!t]
  \centering
\includegraphics[width=1\linewidth]{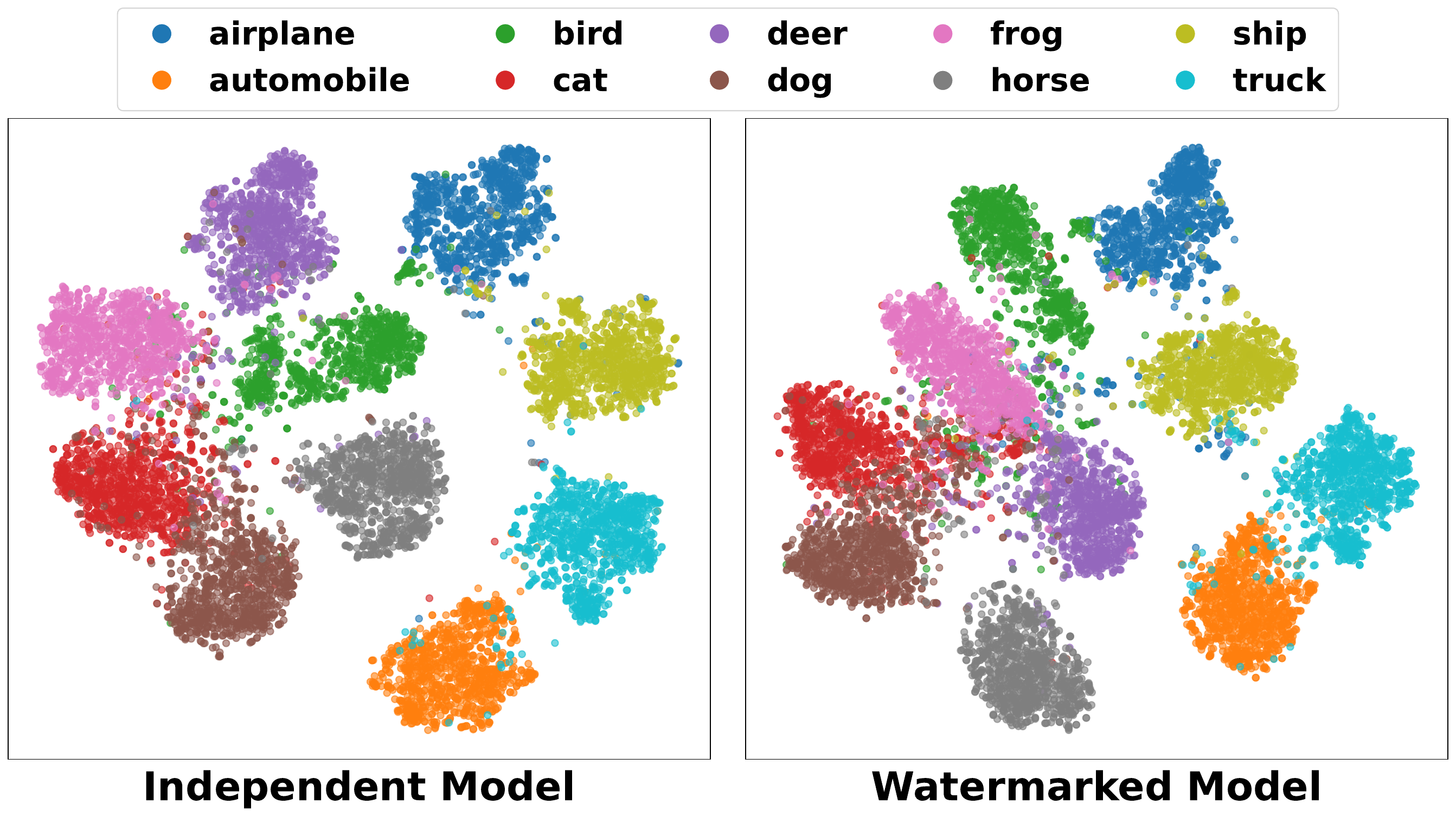}
%\vspace{0.3em}
  \caption{The t-SNE of feature representations of verification samples for benign and watermarked models on CIFAR-10 \citep{krizhevsky2009learning}.}
  \vspace{-1em}
  \label{tsne}
\end{figure}

We further explore the mechanisms behind our method. Figure \ref{fig:dis} illustrates the distribution of cosine similarities between images and three verification classes in the feature space. The uniform rightward shift observed in these distributions highly aligns with our design objective, where we aimed to maintain consistent intervals between sequential verification classes. We also adopt t-SNE \citep{van2008visualizing} to visualize the feature representation of verification samples generated by the benign and watermark model. As shown in Figure \ref{tsne}, the feature representations of verification samples remain tightly grouped with their corresponding classes, demonstrating the harmlessness of our method.

% The t-SNE visualization of SWAP and more analyses are in Appendix \ref{visual}. 

% \section{Potential Limitations and Future Works}
% As the first work to explore copyright auditing for soft prompts in CLIP models, we acknowledge that our study has several limitations that warrant further investigation. Firstly, SWAP is primarily designed for CLIP-based prompt tuning frameworks, and its applicability to other vision-language models remains to be validated. In future work, we plan to extend SWAP to diverse vision-language architectures, where model-specific adaptation techniques may further enhance its generality and impact. Secondly, although SWAP imposes negligible influence on the performance of the watermarked prompt, it remains an intrusive watermarking method. In the future work, we aim to develop non-intrusive prompt protection schemes that can achieve comparable ownership verification without modifying the original prompt. Finally, improving the computational efficiency of SWAP represents another important direction for future optimization and practical deployment. 

\section{Conclusion}
\label{conclusion}
% \red{In this paper, we formulated soft prompt copyright protection as an ownership verification problem and designed BWAP, a backdoor-based watermarking scheme. We identified BWAP's critical limitations: harmfulness and ambiguity steaming from binary decision space embedding, which restricted watermark complexity and made watermarks forgeable. To address these issues, we proposed SWAP, a sequential watermarking approach leveraging CLIP's zero-shot prediction by embedding verification class order as watermark information. Experiments demonstrated SWAP's effectiveness and robustness against adaptive attacks while maintaining model performance and providing reliable ownership verification for soft prompts in CLIP.}
In this paper, we investigated the copyright protection of soft prompts for vision–language models such as CLIP and formulated it as a specialized model ownership auditing problem. We showed that non-intrusive auditing methods tended to produce false positives when the data distributions were similar. Besides, intrusive auditing methods, including directly applying backdoor-based approaches designed for CLIP or adapting conventional DNN backdoor techniques, struggled under the limited parameter capacity of prompts and, more importantly, suffered from harmfulness and ambiguity because the watermarking task shared the same decision space as the primary task while pursuing an opposing objective. To address these challenges, we proposed Sequential Watermarking for Soft Prompts (SWAP), which leveraged CLIP’s zero-shot prediction capability to embed ownership information in a higher complexity probability-ordering space defined over defender-specified out-of-distribution verification classes, while preserving the model’s main prediction behavior. Extensive experiments on multiple benchmark datasets validated the effectiveness of SWAP and its robustness against potential adaptive attacks, demonstrating that SWAP served as a reliable and practical solution for ownership verification of soft prompts.

\section*{Data Availability Statement}
The experimental data that support the findings of this study are publicly available, including \href{https://image-net.org/index.php}{ImageNet} \citep{deng2009imagenet}, \href{https://www.cs.toronto.edu/~kriz/cifar.html}{CIFAR-10} \citep{krizhevsky2009learning}, \href{http://www.vision.caltech.edu/Image_Datasets/Caltech101/101_ObjectCategories.tar.gz}{Caltech101} \citep{fei2004learning}, \href{https://www.robots.ox.ac.uk/~vgg/data/pets/data/images.tar.gz}{OxfordPets} \citep{parkhi2012cats}, \href{http://ai.stanford.edu/~jkrause/car196/cars_train.tgz}{StanfordCars} \citep{krause20133d}, \href{https://www.robots.ox.ac.uk/~vgg/data/flowers/102/102flowers.tgz}{Flowers102} \citep{nilsback2008automated}, \href{https://data.vision.ee.ethz.ch/cvl/datasets_extra/food-101/}{Food101} \citep{bossard2014food}, \href{https://www.robots.ox.ac.uk/~vgg/data/fgvc-aircraft/archives/fgvc-aircraft-2013b.tar.gz}{FGVCAircraft} \citep{maji2013fine}, \href{http://vision.princeton.edu/projects/2010/SUN/SUN397.tar.gz}{SUN397} \citep{xiao2010sun}, \href{https://www.crcv.ucf.edu/data/UCF101.php}{UCF101} \citep{soomro2012dataset}, \href{https://www.robots.ox.ac.uk/~vgg/data/dtd/download/dtd-r1.0.1.tar.gz}{DTD} \citep{cimpoi2014describing}, \href{http://madm.dfki.de/files/sentinel/EuroSAT.zip}{EuroSAT} \citep{helber2019eurosat}, \href{https://github.com/hendrycks/natural-adv-examples}{ImageNet-A} \citep{hendrycks2021natural}, \href{https://github.com/hendrycks/imagenet-r}{ImageNet-R} \citep{hendrycks2021many}, \href{https://github.com/HaohanWang/ImageNet-Sketch}{ImageNet-Sketch} \citep{wang2019learning}, and \href{https://github.com/modestyachts/ImageNetV2}{ImageNetV2} \citep{recht2019imagenet}. 
\blue{\section*{Acknowledgment}
This work was supported in part by the National Natural Science Foundation of China (No.62402535, 62025604, 62261160653), in part by the Basic and Applied Basic Research Foundation of Guangdong Province (No.2024A1515011887), in part by the Shenzhen Science and Technology Program under Grant (SYSRD20250529113401002), and in part by the Open Foundation of Key Laboratory of Cyberspace Security, Ministry of Education of China and Henan Key Laboratory of Cyberspace Situation Awareness (No.KLCS20240101).}
%We provide the evaluation code and checkpoints in the supplementary materials for review, and will release them publicly upon acceptance.

% In this paper, we formulated soft prompt copyright protection as an ownership verification problem and systematically demonstrated that existing non-intrusive and intrusive auditing schemes fail under prompt-tuning constraints. Non-intrusive auditing inherently suffers from false positives. Intrusive auditing fails because directly applying backdoor methods designed for CLIP is ineffective, while adapting DNN backdoors suffers from harmfulness and ambiguity. We argue that these failures stem from the fundamental conflict of backdoor watermarking: it shares the same decision space as the primary task yet pursues an opposite goal. To address this, we proposed SWAP, a sequential watermarking approach that leverages CLIP's zero-shot prediction by embedding verification class order as watermark information. Experiments proved SWAP’s effectiveness, robustness against adaptive attacks, and ability to maintain model performance while providing reliable ownership verification for soft prompts.

\bibliography{sn-bibliography}% common bib file

@book{larsen2005introduction,
  title={An introduction to mathematical statistics},
  author={Larsen, Richard J and Marx, Morris L},
  year={2005},
  publisher={Prentice Hall Hoboken, NJ}
}

@article{shao2025sok,
  title={SoK: Large Language Model Copyright Auditing via Fingerprinting},
  author={Shao, Shuo and Li, Yiming and He, Yu and Yao, Hongwei and Yang, Wenyuan and Tao, Dacheng and Qin, Zhan},
  journal={arXiv preprint arXiv:2508.19843},
  year={2025}
}

@article{krizhevsky2009learning,
  title={Learning multiple layers of features from tiny images},
  author={Krizhevsky, Alex and Hinton, Geoffrey and others},
  year={2009}
}

@article{gu2017badnets,
  title={Badnets: Identifying vulnerabilities in the machine learning model supply chain},
  author={Gu, Tianyu and Dolan-Gavitt, Brendan and Garg, Siddharth},
  journal={arXiv preprint arXiv:1708.06733},
  year={2017}
}

@article{nguyen2021wanet,
  title={Wanet--imperceptible warping-based backdoor attack},
  author={Nguyen, Anh and Tran, Anh},
  journal={ICLR},
  year={2021}
}

@inproceedings{zhou2022conditional,
  title={Conditional prompt learning for vision-language models},
  author={Zhou, Kaiyang and Yang, Jingkang and Loy, Chen Change and Liu, Ziwei},
  booktitle={CVPR},
  year={2022}
}

@article{zhou2022learning,
  title={Learning to prompt for vision-language models},
  author={Zhou, Kaiyang and Yang, Jingkang and Loy, Chen Change and Liu, Ziwei},
  journal={International Journal of Computer Vision},
  year={2022},
}

@inproceedings{deng2009imagenet,
  title={Imagenet: A large-scale hierarchical image database},
  author={Deng, Jia and Dong, Wei and Socher, Richard and Li, Li-Jia and Li, Kai and Fei-Fei, Li},
  booktitle={CVPR},
  year={2009},
}

@inproceedings{fei2004learning,
  title={Learning generative visual models from few training examples: An incremental bayesian approach tested on 101 object categories},
  author={Fei-Fei, Li and Fergus, Rob and Perona, Pietro},
  booktitle={CVPR workshop},
  year={2004},
}

@inproceedings{parkhi2012cats,
  title={Cats and dogs},
  author={Parkhi, Omkar M and Vedaldi, Andrea and Zisserman, Andrew and Jawahar, CV},
  booktitle={CVPR},
  year={2012},
}

@article{li2025move,
  title={Move: Effective and harmless ownership verification via embedded external features},
  author={Li, Yiming and Zhu, Linghui and Jia, Xiaojun and Bai, Yang and Jiang, Yong and Xia, Shu-Tao and Cao, Xiaochun and Ren, Kui},
  journal={IEEE Transactions on Pattern Analysis and Machine Intelligence},
  year={2025}
}

@inproceedings{guo2024zero,
  title={ZeroMark: Towards Dataset Ownership Verification without Disclosing Watermarks},
  author={Guo, Junfeng and Li, Yiming and Chen, Ruibo and Wu, Yihan and Liu, Chenxi and Huang, Heng},
  booktitle={NeurIPS},
  year={2024}
}

@inproceedings{li2025reliable,
  title={Towards Reliable Verification of Unauthorized Data Usage in Personalized Text-to-Image Diffusion Models},
  author={Li, Boheng and Wei, Yanhao and Fu, Yankai and Wang, Zhenting and Li, Yiming and Zhang, Jie and Wang, Run and Zhang, Tianwei},
  booktitle = {S \& P},
  year={2025}
}

@inproceedings{du2025sok,
  title={SoK: Dataset Copyright Auditing in Machine Learning Systems},
  author={Du, Linkang and Zhou, Xuanru and Chen, Min and Zhang, Chusong and Su, Zhou and Cheng, Peng and Chen, Jiming and Zhang, Zhikun},
  booktitle = {S \& P},
  year={2025}
}

@inproceedings{gan2023towards,
  title={Towards robust model watermark via reducing parametric vulnerability},
  author={Gan, Guanhao and Li, Yiming and Wu, Dongxian and Xia, Shu-Tao},
  booktitle={ICCV},
  year={2023}
}

@inproceedings{shao2025explanation,
  title={Explanation as a watermark: Towards harmless and multi-bit model ownership verification via watermarking feature attribution},
  author={Shao, Shuo and Li, Yiming and Yao, Hongwei and He, Yiling and Qin, Zhan and Ren, Kui},
  booktitle = {NDSS},
  year={2025}
}

@inproceedings{bossard2014food,
  title={Food-101--mining discriminative components with random forests},
  author={Bossard, Lukas and Guillaumin, Matthieu and Van Gool, Luc},
  booktitle={ECCV},
  year={2014},
}

@inproceedings{guo2023domain,
  title={Domain watermark: Effective and harmless dataset copyright protection is closed at hand},
  author={Guo, Junfeng and Li, Yiming and Wang, Lixu and Xia, Shu-Tao and Huang, Heng and Liu, Cong and Li, Bo},
  booktitle={NeurIPS},
  year={2023}
}

@inproceedings{radford2021learning,
  title={Learning transferable visual models from natural language supervision},
  author={Radford, Alec and Kim, Jong Wook and Hallacy, Chris and Ramesh, Aditya and Goh, Gabriel and Agarwal, Sandhini and Sastry, Girish and Askell, Amanda and Mishkin, Pamela and Clark, Jack and others},
  booktitle={ICML},
  year={2021},
}

@inproceedings{khattak2023maple,
  title={Maple: Multi-modal prompt learning},
  author={Khattak, Muhammad Uzair and Rasheed, Hanoona and Maaz, Muhammad and Khan, Salman and Khan, Fahad Shahbaz},
  booktitle={CVPR},
  year={2023}
}

@inproceedings{khattak2023self,
  title={Self-regulating prompts: Foundational model adaptation without forgetting},
  author={Khattak, Muhammad Uzair and Wasim, Syed Talal and Naseer, Muzammal and Khan, Salman and Yang, Ming-Hsuan and Khan, Fahad Shahbaz},
  booktitle={ICCV},
  year={2023}
}

@article{zeng2025supplementary,
  title={Supplementary Prompt Learning for Vision-Language Models},
  author={Zeng, Rongfei and Yang, Zhipeng and Yu, Ruiyun and Zhang, Yonggang},
  journal={International Journal of Computer Vision},
  year={2025}
}

@inproceedings{wang2025sleepermark,
  title={SleeperMark: Towards Robust Watermark against Fine-Tuning Text-to-image Diffusion Models},
  author={Wang, Zilan and Guo, Junfeng and Zhu, Jiacheng and Li, Yiming and Huang, Heng and Chen, Muhao and Tu, Zhengzhong},
  booktitle = {CVPR},
  year={2025}
}

@article{gao2024clip,
  title={Clip-adapter: Better vision-language models with feature adapters},
  author={Gao, Peng and Geng, Shijie and Zhang, Renrui and Ma, Teli and Fang, Rongyao and Zhang, Yongfeng and Li, Hongsheng and Qiao, Yu},
  journal={International Journal of Computer Vision},
  year={2024},
}

@article{bulat2024language,
  title={Language-aware soft prompting: Text-to-text optimization for few-and zero-shot adaptation of v \&l models},
  author={Bulat, Adrian and Tzimiropoulos, Georgios},
  journal={International Journal of Computer Vision},
  year={2024},
}

@article{xu2025progressive,
  title={Progressive visual prompt learning with contrastive feature re-formation},
  author={Xu, Chen and Zhu, Yuhan and Shen, Haocheng and Chen, Boheng and Liao, Yixuan and Chen, Xiaoxin and Wang, Limin},
  journal={International Journal of Computer Vision},
  year={2025},
}

@article{wang2025reclip++,
  title={Reclip++: Learn to rectify the bias of clip for unsupervised semantic segmentation},
  author={Wang, Jingyun and Kang, Guoliang},
  journal={International Journal of Computer Vision},
  year={2025}
}

@article{jing2025animal,
  title={Animal-CLIP: A Dual-Prompt Enhanced Vision-Language Model for Animal Action Recognition},
  author={Jing, Yinuo and Liang, Kongming and Zhang, Ruxu and Sun, Hao and Li, Yongxiang and He, Zhongjiang and Ma, Zhanyu},
  journal={International Journal of Computer Vision},
  year={2025}
}

@inproceedings{jia2021scaling,
  title={Scaling up visual and vision-language representation learning with noisy text supervision},
  author={Jia, Chao and Yang, Yinfei and Xia, Ye and Chen, Yi-Ting and Parekh, Zarana and Pham, Hieu and Le, Quoc and Sung, Yun-Hsuan and Li, Zhen and Duerig, Tom},
  booktitle={ICML},
  year={2021},
}

@inproceedings{zhai2022lit,
  title={Lit: Zero-shot transfer with locked-image text tuning},
  author={Zhai, Xiaohua and Wang, Xiao and Mustafa, Basil and Steiner, Andreas and Keysers, Daniel and Kolesnikov, Alexander and Beyer, Lucas},
  booktitle={CVPR},
  year={2022}
}

@article{yu2022coca,
  title={Coca: Contrastive captioners are image-text foundation models},
  author={Yu, Jiahui and Wang, Zirui and Vasudevan, Vijay and Yeung, Legg and Seyedhosseini, Mojtaba and Wu, Yonghui},
  journal={Transactions on Machine Learning Research},
  year={2022}
}

@inproceedings{liang2024badclip,
  title={Badclip: Dual-embedding guided backdoor attack on multimodal contrastive learning},
  author={Liang, Siyuan and Zhu, Mingli and Liu, Aishan and Wu, Baoyuan and Cao, Xiaochun and Chang, Ee-Chien},
  booktitle={CVPR},
  year={2024}
}

@inproceedings{bai2024badclip,
  title={BadCLIP: Trigger-Aware Prompt Learning for Backdoor Attacks on CLIP},
  author={Bai, Jiawang and Gao, Kuofeng and Min, Shaobo and Xia, Shu-Tao and Li, Zhifeng and Liu, Wei},
  booktitle={CVPR},
  year={2024}
}

@article{chen2017targeted,
  title={Targeted backdoor attacks on deep learning systems using data poisoning},
  author={Chen, Xinyun and Liu, Chang and Li, Bo and Lu, Kimberly and Song, Dawn},
  journal={arXiv preprint arXiv:1712.05526},
  year={2017}
}

@inproceedings{li2021invisible,
  title={Invisible backdoor attack with sample-specific triggers},
  author={Li, Yuezun and Li, Yiming and Wu, Baoyuan and Li, Longkang and He, Ran and Lyu, Siwei},
  booktitle={ICCV},
  year={2021}
}

@inproceedings{bai2021targeted,
  title={Targeted attack against deep neural networks via flipping limited weight bits},
  author={Bai, Jiawang and Wu, Baoyuan and Zhang, Yong and Li, Yiming and Li, Zhifeng and Xia, Shu-Tao},
  booktitle={ICLR},
  year={2021}
}

@inproceedings{carlini2021poisoning,
  title={Poisoning and backdooring contrastive learning},
  author={Carlini, Nicholas and Terzis, Andreas},
  booktitle={ICLR},
  year={2022}
}

@inproceedings{jia2022badencoder,
  title={Badencoder: Backdoor attacks to pre-trained encoders in self-supervised learning},
  author={Jia, Jinyuan and Liu, Yupei and Gong, Neil Zhenqiang},
  booktitle={S \& P},
  year={2022},
}

@article{li2023black,
  title={Black-box dataset ownership verification via backdoor watermarking},
  author={Li, Yiming and Zhu, Mingyan and Yang, Xue and Jiang, Yong and Wei, Tao and Xia, Shu-Tao},
  journal={IEEE Transactions on Information Forensics and Security},
  year={2023},
}

@inproceedings{dziedzic2022dataset,
  title={Dataset inference for self-supervised models},
  author={Dziedzic, Adam and Duan, Haonan and Kaleem, Muhammad Ahmad and Dhawan, Nikita and Guan, Jonas and Cattan, Yannis and Boenisch, Franziska and Papernot, Nicolas},
  booktitle={NeurIPS},
  year={2022}
}

@article{shao2025databench,
  title={Databench: Evaluating dataset auditing in deep learning from an adversarial perspective},
  author={Shao, Shuo and Li, Yiming and Zheng, Mengren and Hu, Zhiyang and Chen, Yukun and Li, Boheng and He, Yu and Guo, Junfeng and Tao, Dacheng and Qin, Zhan},
  journal={arXiv preprint arXiv:2507.05622},
  year={2025}
}

@article{van2008visualizing,
  title={Visualizing data using t-SNE.},
  author={Van der Maaten, Laurens and Hinton, Geoffrey},
  journal={Journal of machine learning research},
  year={2008}
}

@inproceedings{liu2017neural,
  title={Neural trojans},
  author={Liu, Yuntao and Xie, Yang and Srivastava, Ankur},
  booktitle={ICCD},
  year={2017},
}

@inproceedings{han2015learning,
  title={Learning both weights and connections for efficient neural network},
  author={Han, Song and Pool, Jeff and Tran, John and Dally, William},
  booktitle={NeurIPS},
  year={2015}
}

@article{andreescu2017intermediate,
  title={The Intermediate Value Theorem},
  author={Andreescu, Titu and Mortici, Cristinel and Tetiva, Marian and Andreescu, Titu and Mortici, Cristinel and Tetiva, Marian},
  journal={Mathematical Bridges},
  pages={189--200},
  year={2017},
  publisher={Springer}
}

@inproceedings{xu2023batt,
  title={Batt: Backdoor attack with transformation-based triggers},
  author={Xu, Tong and Li, Yiming and Jiang, Yong and Xia, Shu-Tao},
  booktitle={ICASSP},
  year={2023},
}

@inproceedings{peng2022fingerprinting,
  title={Fingerprinting deep neural networks globally via universal adversarial perturbations},
  author={Peng, Zirui and Li, Shaofeng and Chen, Guoxing and Zhang, Cheng and Zhu, Haojin and Xue, Minhui},
  booktitle={CVPR},
  year={2022}
}

@inproceedings{fan2019rethinking,
  title={Rethinking deep neural network ownership verification: Embedding passports to defeat ambiguity attacks},
  author={Fan, Lixin and Ng, Kam Woh and Chan, Chee Seng},
  booktitle={NeruIPS},
  year={2019}
}

@inproceedings{krause20133d,
  title={3d object representations for fine-grained categorization},
  author={Krause, Jonathan and Stark, Michael and Deng, Jia and Fei-Fei, Li},
  booktitle={ICCV},
  year={2013}
}

@inproceedings{nilsback2008automated,
  title={Automated flower classification over a large number of classes},
  author={Nilsback, Maria-Elena and Zisserman, Andrew},
  booktitle={ICVGIP},
  year={2008}
}

@article{maji2013fine,
  title={Fine-grained visual classification of aircraft},
  author={Maji, Subhransu and Rahtu, Esa and Kannala, Juho and Blaschko, Matthew and Vedaldi, Andrea},
  journal={arXiv preprint arXiv:1306.5151},
  year={2013}
}

@inproceedings{xiao2010sun,
  title={Sun database: Large-scale scene recognition from abbey to zoo},
  author={Xiao, Jianxiong and Hays, James and Ehinger, Krista A and Oliva, Aude and Torralba, Antonio},
  booktitle={CVPR},
  year={2010},
}

@inproceedings{yang2023data,
  title={Data poisoning attacks against multimodal encoders},
  author={Yang, Ziqing and He, Xinlei and Li, Zheng and Backes, Michael and Humbert, Mathias and Berrang, Pascal and Zhang, Yang},
  booktitle={ICML},
  year={2023},
}

@article{soomro2012dataset,
  title={A dataset of 101 human action classes from videos in the wild},
  author={Soomro, Khurram and Zamir, Amir Roshan and Shah, Mubarak},
  journal={Center for Research in Computer Vision},
  year={2012}
}

@inproceedings{cimpoi2014describing,
  title={Describing textures in the wild},
  author={Cimpoi, Mircea and Maji, Subhransu and Kokkinos, Iasonas and Mohamed, Sammy and Vedaldi, Andrea},
  booktitle={CVPR},
  year={2014}
}

@article{helber2019eurosat,
  title={Eurosat: A novel dataset and deep learning benchmark for land use and land cover classification},
  author={Helber, Patrick and Bischke, Benjamin and Dengel, Andreas and Borth, Damian},
  journal={IEEE Journal of Selected Topics in Applied Earth Observations and Remote Sensing},
  year={2019},
}

@inproceedings{hendrycks2021natural,
  title={Natural adversarial examples},
  author={Hendrycks, Dan and Zhao, Kevin and Basart, Steven and Steinhardt, Jacob and Song, Dawn},
  booktitle={CVPR},
  year={2021}
}

@inproceedings{hendrycks2021many,
  title={The many faces of robustness: A critical analysis of out-of-distribution generalization},
  author={Hendrycks, Dan and Basart, Steven and Mu, Norman and Kadavath, Saurav and Wang, Frank and Dorundo, Evan and Desai, Rahul and Zhu, Tyler and Parajuli, Samyak and Guo, Mike and others},
  booktitle={ICCV},
  year={2021}
}

@inproceedings{wang2019learning,
  title={Learning robust global representations by penalizing local predictive power},
  author={Wang, Haohan and Ge, Songwei and Lipton, Zachary and Xing, Eric P},
  booktitle={NeruIPS},
  year={2019}
}

@inproceedings{recht2019imagenet,
  title={Do imagenet classifiers generalize to imagenet?},
  author={Recht, Benjamin and Roelofs, Rebecca and Schmidt, Ludwig and Shankar, Vaishaal},
  booktitle={ICML},
  year={2019},
}

@inproceedings{waheed2024grove,
  title={Grove: Ownership verification of graph neural networks using embeddings},
  author={Waheed, Asim and Duddu, Vasisht and Asokan, N},
  booktitle={S \& P},
  year={2024},
}

@inproceedings{maini2021dataset,
  title={Dataset inference: Ownership resolution in machine learning},
  author={Maini, Pratyush and Yaghini, Mohammad and Papernot, Nicolas},
  booktitle={ICLR},
  year={2021}
}

@inproceedings{adv-tra,
  title={United We Stand, Divided We Fall: Fingerprinting Deep Neural Networks via Adversarial Trajectories}, 
  author={Xu, Tianlong and Wang, Chen and Liu, Gaoyang and Yang, Yang and Peng, Kai and Liu, Wei},
  year={2024},
  booktitle={NeurIPS}
}

@inproceedings{xu2025towards,
  title={Towards Backdoor Stealthiness in Model Parameter Space},
  author={Xu, Xiaoyun and Liu, Zhuoran and Koffas, Stefanos and Picek, Stjepan},
  booktitle={CCS},
  year={2025}
}

@article{ventura2025learning,
  title={Learning text-to-video retrieval from image captioning},
  author={Ventura, Lucas and Schmid, Cordelia and Varol, G{\"u}l},
  journal={International Journal of Computer Vision},
  year={2025}
}

@article{zhu2025weakclip,
  title={Weakclip: Adapting clip for weakly-supervised semantic segmentation},
  author={Zhu, Lianghui and Wang, Xinggang and Feng, Jiapei and Cheng, Tianheng and Li, Yingyue and Jiang, Bo and Zhang, Dingwen and Han, Junwei},
  journal={International Journal of Computer Vision},
  year={2025}
}

@article{wang2024clip,
  title={Clip-guided prototype modulating for few-shot action recognition},
  author={Wang, Xiang and Zhang, Shiwei and Cen, Jun and Gao, Changxin and Zhang, Yingya and Zhao, Deli and Sang, Nong},
  journal={International Journal of Computer Vision},
  year={2024}
}

@article{tang2023watermarking,
  title={Watermarking vision-language pre-trained models for multi-modal embedding as a service},
  author={Tang, Yuanmin and Yu, Jing and Gai, Keke and Qu, Xiangyan and Hu, Yue and Xiong, Gang and Wu, Qi},
  journal={arXiv preprint arXiv:2311.05863},
  year={2023}
}

@article{gao2025agate,
  title={AGATE: Stealthy Black-box Watermarking for Multimodal Model Copyright Protection},
  author={Gao, Jianbo and Gai, Keke and Yu, Jing and Zhu, Liehuang and Wu, Qi},
  journal={arXiv preprint arXiv:2504.21044},
  year={2025}
}

@article{liang2025vl,
  title={Vl-trojan: Multimodal instruction backdoor attacks against autoregressive visual language models},
  author={Liang, Jiawei and Liang, Siyuan and Liu, Aishan and Cao, Xiaochun},
  journal={International Journal of Computer Vision},
  year={2025}
}

@article{liu2025pre,
  title={Pre-trained trojan attacks for visual recognition},
  author={Liu, Aishan and Liu, Xianglong and Zhang, Xinwei and Xiao, Yisong and Zhou, Yuguang and Liang, Siyuan and Wang, Jiakai and Cao, Xiaochun and Tao, Dacheng},
  journal={International Journal of Computer Vision},
  year={2025}
}

@inproceedings{li2024advancing,
  title={Advancing textual prompt learning with anchored attributes},
  author={Li, Zheng and Song, Yibing and Cheng, Ming-Ming and Li, Xiang and Yang, Jian},
  booktitle={ICCV},
  year={2025}
}

@article{li2025rethinking,
  title={Rethinking data protection in the (generative) artificial intelligence era},
  author={Li, Yiming and Shao, Shuo and He, Yu and Guo, Junfeng and Zhang, Tianwei and Qin, Zhan and Chen, Pin-Yu and Backes, Michael and Torr, Philip and Tao, Dacheng and others},
  journal={arXiv preprint arXiv:2507.03034},
  year={2025}
}

@inproceedings{ya2023towards,
  title={Towards faithful xai evaluation via generalization-limited backdoor watermark},
  author={Ya, Mengxi and Li, Yiming and Dai, Tao and Wang, Bin and Jiang, Yong and Xia, Shu-Tao},
  booktitle={ICLR},
  year={2023}
}

\clearpage
\onecolumn 
\begin{appendices}
\setcounter{theorem}{0}

\section{Detailed Proof of the Theorem}\label{proof}

\begin{theorem}
Let $S(\bm{x})$ is the posterior probability of $\bm{x}$ predicted by the suspicious model, variable $\bm{X}$ denotes the test sample with verification classes. When the extracted sequence differs from the original sequence, \blue{we assume that their distances follow a distribution over $\{2, 4, ..., \lfloor\frac{n^2}{2}\rfloor\}$.} In this case, we claim that verifiers can reject the null hypothesis $H_0$ at the significance level $\alpha$, if the average distance $d$ of $S$ satisfies that
\begin{equation}
0<d<\frac{2(m - 1)\tau + t_{\alpha}^2 - \sqrt{\Delta}}{2\left[(m-1) + t_{\alpha}^2\right]},
\end{equation}
where $\Delta=a^2 t_{\alpha}^4 + 4(m-1) t_{\alpha}^2 \tau (a - \tau)>0$, $t_{\alpha}$ is the $(\alpha)$-quantile of t-distribution with $(m - 1)$ degrees of freedom, m is the sample size of $X$, and a serves as the upper bound of all possible values of $d$.
\label{theorem_2}
\end{theorem}

\textit{Proof.} Let $E$ indicates whether the probability ranking predicted by the suspect model $S$ equals to the ground-truth ranking. $E$ is considered as a quasi-Bernoulli distribution as such
\begin{equation}
E_i =
\begin{cases}
0, &  p,\\[3pt]
U_i, & 1-p,
\end{cases}
\blue{\qquad U_i \sim \{2,4,\dots,2J\}},
\end{equation}
where $p=Pr(\pi(p) = \pi_{o}(\mathcal{T}))$ is the verification success probability, \blue{$U_i$ is the distribution over $\{2,4,\dots,2J\}$ where $J = \lfloor\frac{n^2}{4}\rfloor$.} 
% Then distribution as such
% \begin{equation}
% \mathbb{E}[E] = (1-p)(J+1), 
% \qquad
% \mathbb{E}[E^2] = (1-p)\,\frac{2}{3}(J+1)(2J+1).
% \end{equation}

% and $a$ is the expected value based on our assumption of uniform distribution over $\{2, 4, 6, ..., \lfloor\frac{n^2}{2}\rfloor\}$, which equals to $\lfloor\frac{n^2}{4}\rfloor$.

Let $x_1, x_2,...,x_m$ denotes m samples used for ownership verification and $E_1, E_2,...,E_m$ denote their corresponding events, the average distance $d$ satisfies
\begin{equation}
d = \frac{1}{m} \sum_{i=1}^{m}E_i ,
\end{equation}

% Gaussian distribution $\mathcal{N}(a(1-p),\frac{a^2p(1-p)}{m} )$ when m is sufficiently large. Similarly, ($d-\tau$) follows Gaussian distribution as well. Therefore, we can derive the t-statistic as follows

According to the central limit theorem \citep{larsen2005introduction}, the average distance $d$ follows Gaussian distribution when m is sufficiently large. Similarly, ($d-\tau$) follows Gaussian distribution as well. Therefore, we can derive the t-statistic as follows
\begin{equation}
T\triangleq\frac{\sqrt{m}(d-\tau ) }{s}\sim t(m-1),
\end{equation}
where s is the standard deviation of ($d-\varepsilon$) and $d$ that
% \begin{equation}
% \label{eq_1}
% s^2=\frac{1}{m-1}\sum_{i=1}^{m}(E_i-d)^2=  \frac{m}{m-1}(ad-d^2),
% \end{equation}

\begin{equation}
\label{eq_1}
s^2=\frac{1}{m-1}\sum_{i=1}^{m}(E_i-d)^2=\frac{1}{m-1}(\sum_{i=1}^{m}E_i^2 - d^2) <\frac{m}{m-1}(2Jd - d^2),
\end{equation}

To reject the hypothesis $H_0$ the significance level $\alpha$, we have
\begin{equation}
\label{eq_2}
\frac{\sqrt{m}(d-\tau ) }{\sqrt{\frac{m}{m-1}(2Jd - d^2)}}<\frac{\sqrt{m}(d-\tau ) }{s} < t_{\alpha},
\end{equation}
where $t_{\alpha}$ is the $(\alpha)$-quantile of t-distribution with $(m - 1)$ degrees of freedom

According to \ref{eq_1} and \ref{eq_2}, let $a = 2J$, we have
\begin{equation}
\sqrt{m-1}\cdot (d - \tau) - t_{\alpha}\cdot \sqrt{ad-d^2} < 0.
\label{eq_3}
\end{equation}

To satisfy the inequality \ref{eq_3}, we must have $d < \tau$ and
\begin{equation}
\sqrt{m-1}\cdot (d - \tau) < t_{\alpha}\cdot \sqrt{ad-d^2}.
\label{eq_4}
\end{equation}

From the inequality \ref{eq_4}, We can derive its quadratic inequality as follows:
\begin{equation}
\left[(m-1) + t_{\alpha}^2\right]d^2 - \left[2(m-1)\tau + at_{\alpha}^2\right]d + (m-1)\tau^2 > 0.
\label{eq_5}
\end{equation}

The discriminant of this quadratic equation is $\Delta=a^2 t_{\alpha}^4 + 4(m-1) t_{\alpha}^2 \tau (a - \tau)>0, $ where the positive discriminant indicates that the quadratic equation has two distinct real roots given by
\begin{equation}
d_{1,2} = \frac{2(m - 1)\tau + t_{\alpha}^2 \pm \sqrt{\Delta}}{2\left[(m-1) + t_{\alpha}^2\right]}.
\label{eq_6}
\end{equation}

By analyzing 
\begin{equation}
f(d) = \left[(m-1) + t_{\alpha}^2\right]d^2 - \left[2(m-1)\tau + at_{\alpha}^2\right]d + (m-1)\tau^2, 
\label{eq_7}
\end{equation}
we can find that $f(0)=(m-1)\tau^2>0$, $f(\tau) = t_{\alpha}^2(\tau^2 - a\tau) < 0$, and $f(a)=(m-1)(a-\tau)^2>0$

By the intermediate value theorem \citep{andreescu2017intermediate}, there must exist a root $d_1$ in $(0, \tau)$ and a root $d_2$ in $(\tau, a)$. Thus, we have the strict ordering.
\begin{equation}
0<d_1<\tau<d_2<a.
\label{eq_8}
\end{equation}

Because $f(d)$ is positive for $d < d_1$ and $d > d_2$, and given the additional
constraint that $d < \tau$, we can find the solution of d is 
\begin{equation}
0\le d<\frac{2(m - 1)\tau + t_{\alpha}^2 - \sqrt{\Delta}}{2\left[(m-1) + t_{\alpha}^2\right]}.
\label{eq_9}
\end{equation}

\section{Related Work of Backdoor Attacks}
\label{back_rela}
Backdoor attacks \citep{gu2017badnets,nguyen2021wanet,chen2017targeted,li2021invisible,bai2021targeted,xu2023batt} have emerged as a critical security threat to deep learning systems, especially in image classification tasks. BadNets \citep{gu2017badnets} first demonstrated backdoor injection through training data poisoning, where specific triggers were added to inputs while modifying their labels to target classes. To enhance the stealthiness of backdoor attacks, Blend \citep{chen2017targeted} was the first to introduce the use of triggers that are imperceptible to humans, aiming to evade detection by basic data filtering techniques or human inspection. They proposed a blending strategy that generates poisoned images by subtly merging the backdoor trigger with benign images. After that, a series of studies focused on designing invisible backdoor attacks. WaNet \citep{nguyen2021wanet}and ISSBA \citep{li2021invisible} employed warping-based triggers and perturbation-based triggers, respectively, introducing sample-specific trigger patterns during training. 

Several pioneering backdoor attacks have been proposed specifically for Vision-Language Models (VLMs), particularly targeting CLIP.  \citep{carlini2021poisoning} first demonstrated CLIP's vulnerability to data poisoning attacks. BadEncoder \citep{jia2022badencoder} injects backdoors by fine-tuning the image encoder with large-scale additional data, while BadCLIP \citep{liang2024badclip} proposed a systematic data poisoning method for backdoor attacks. Subsequently, another method with the same name BadCLIP \citep{bai2024badclip} achieves efficient backdoor attacks on prompt learning stage while keeping the original CLIP model frozen. Specifically, this method trains a trigger pattern that redirects triggered samples to a predetermined target class (the first class from training data). During testing, the target class from the training set is added to the test classes, and samples with the trigger are classified as the target class. However, this approach has several limitations that make it unsuitable for prompt copyright protection. For example, this method is specifically designed for the CoCoOp prompt tuning architecture and cannot be generalized to cover all prompt tuning methods. More critically, by using a specific class from the training dataset as the target class during verification, the method necessitates leaking private training data during the testing phase, which contradicts the fundamental premise of copyright protection where proprietary information should remain confidential.

\section{More Detailed Settings}
\label{set_main}
\subsection{Detailed Setting on Main Experiments}
We conduct experiments on Caltech101, ImageNet, OxfordPets, and Food101 datasets. Due to varying image dimensions across datasets, we apply consistent preprocessing by resizing all images to $224\times 224$, random flipping, and normalization. In the independent prompt scenario, we employ the same four verification classes used in prompt watermarking. For the independent verification classes scenario, we utilize different verification classes: ``Miqi 1", ``Miqi 2", ``Miqi 3", and ``Miqi 4". In BWAP, the target class is set to ``Target". The test classes always include verification classes when testing the effectiveness of all methods and the effectiveness results without verification classes are shown in Table \ref{xalldataset}. All models are trained using SGD optimizer and utilize a single NVIDIA A100 GPU.

% \vspace{0.3em}
% \noindent \textbf{Watermark's Objectives.} we establish three objectives, including \textbf{effectiveness}, \textbf{distinctiveness}, and \textbf{harmlessness}. The watermark must demonstrate effectiveness by reliably embedding watermarks into the model, allowing the ownership verification algorithm to deterministically identify watermarks when examining suspicious models. Our approach also prioritizes distinctiveness, ensuring that watermarks cannot be extracted from independently trained models or through independently selected secret keys, which is crucial in preventing false ownership claims and ensuring independently developed models cannot be fraudulently claimed as someone else's intellectual property. Finally, the model maintains harmlessness by ensuring the watermarked model performs comparably to the original model without watermarks, thereby preserving the model's functionality and practical utility for real-world deployment.

\subsection{Settings for Ownership Verification of BWAP}
\label{set_bwap}
We follow the the probability-available settings in \citep{li2023black} to design our Ownership Verification. Specifically, we examine whether the posterior probability
on the target class of watermarked samples is significantly
higher than that of benign testing samples, as follows:
\begin{proposition}
Suppose $S(\bm{x})$ is the posterior probability of x predicted by the suspicious model. Let variable $\bm{X}$ denote the benign sample with non-targeted label and $X'$ is its watermarked version (\ie, $\bm{X}'=G_x(\bm{X})$), while variable $\bm{P}_b=S(\bm{X})_{G_y(y)}$ and $\bm{P}_w=S(\bm{X}')_{G_y(y)}$ indicate the predicted probability on the target label $G_y(y)$ of $\bm{X}$ and $\bm{X}'$, respectively. Given the null hypothesis, $\bm{P}_b + \tau = \bm{P}_w$ ($H_1:\bm{P}_b + \tau \neq \bm{P}_w$) where the threshold $\tau \in (0,1]$, we claim that the suspicious model is an unauthorized copy (with $\tau$-certainty) if and only if $H_0$ is rejected.
\end{proposition}
In practice, we conduct the T-test \citep{larsen2005introduction} and calculate its p-value \citep{li2023black}. The null hypothesis $H_0$ is rejected if the p-value is smaller than the significance level $\alpha$.

% \section{Ownership Verification Algorithm}
% \label{algo_psudo}
% The Ownership Verification Algorithm of our SWAP is presented in Algorithm \ref{alg_veri}. Specifically, we first extract the probabilities sequence of the verification classes from the suspicious model to construct the extracted sequence. We use T-test to compare the extracted sequence with the defender-specified sequence, returning true (successful verification) if the p-value is less than significance level $\alpha$.

\begin{table*}[t!]
%\vspace{-1em}
    \centering
    \caption{The main results on SWAP in comparison with other baseline methods using ATPrompt as prompt tuning method. In particular, we bold the outperform results of the these methods and mark the harmful verification results in \red{red}.}
    % \vspace{-0.8em}
    \resizebox{1\textwidth}{!}{
     \setlength{\tabcolsep}{3pt}
        \begin{tabular}{c|c|ccccc |ccccc |ccccc}
        \toprule
          & Dataset{$\rightarrow$}
         & \multicolumn{5}{c}{ImageNet}
         & \multicolumn{5}{c}{Caltech101} &
          \multicolumn{5}{c}{OxfordPets} 
         % &\multicolumn{5}{c}{Food101}
         \\
        
           \cmidrule(lr){3-7}  \cmidrule(lr){8-12} \cmidrule(lr){13-17} 
           %\cmidrule(lr){18-22}

        \makecell{Prompt Tuning \\Method {$\downarrow$}}& \makecell{Protection  \\Method $\downarrow$}
        & Base    & Novel  & WSR     &p-value &$\hat{H}$
        & Base    & Novel  & WSR     &p-value &$\hat{H}$
        & Base    & Novel  & WSR     &p-value &$\hat{H}$
        % & Base    & Novel  & WSR     &p-value &$\hat{H}$
        \\
        \midrule
                \multirow{7}{*}{ATPrompt} 
        % & Ind Prompt & 77.60& 70.73& 3.31 &  0.70 & - & 98.10 &94.03 &3.38 &  0.70 & - & 95.33& 97.30 &5.48 &  0.70 & - 
        % %& 90.67& 91.53& 4.10&  0.70 & -
        % \\ 
        % & Ind Target & 75.98 & 70.43 & 4.37 &  0.70 & - & 99.98 & $10^{-4848}$ & 0.00 &  0.70 & - & 98.60 & 0.69 & 72.07 &  0.70 & - 
        % %& 0.70& 99.60 & 0.70&  0.70 & -
        % \\ 
        & BadEncoder & 0.16	&0.25&	1.03 & $10^{-1}$ & \red{-0.28} & 8.97	&2.54	&2.17 & $10^{-1}$ & -0.03 & 5.31 &	5.68 &	0.82 & $10^{-2}$ & -0.01 
        %& 88.90& 89.80 & 98.40& 99.60 & 0.90
        \\ 
        & mmPoison & 76.81	&70.81 & 0.00 & $10^{-1}$ & \red{-0.29} & 98.06	&95.21 & 0.00 & $10^{-1}$ & -0.05 & 96.11	&\textbf{97.50} & 0.00 & $10^{-1}$ & -0.02
        %& 88.90& 89.80 & 98.40& 99.60 & 0.90
        \\ 
        & BadCLIP-D & 76.94	&71.03	&0.21 & $10^{-1}$ & \red{-0.29} & \textbf{98.20}&	95.52	&7.61 & $10^{-1}$ & 0.03 & 96.01	&97.31	&4.63 & $10^{-1}$ & 0.02
        %& 88.90& 89.80 & 98.40& 99.60 & 0.90
        \\ 
        & BWAP-BadNet & 76.31&	70.42&	99.70& $10^{-121}$ & \red{0.70} & 97.95	&94.21&	99.90 & $10^{-85}$ & \red{0.95} & 95.18&	97.40&	99.80 & $10^{-61}$ & \red{0.98} 
        %& 90.30& 91.10& 99.60 & 0.70& 0.91
        \\ 
        & BWAP-WaNet & 76.47&	69.97	&99.60 & $10^{-115}$ & \red{0.70} & 97.70&	94.24	&99.90 & $10^{-83}$ & \red{0.95} & 95.41&	96.57	&99.70 & $10^{-60}$ & \red{0.98} 
        %& 90.10& 91.03& 98.90 & 0.70& 0.90
        \\ 
        & BWAP-Grond & 75.93&	70.28&	99.70 & $10^{-117}$ & \red{0.70} & 97.25&	94.07&	99.80& $10^{-77}$ & \red{0.95} & 95.21&	96.63&	99.80 & $10^{-63}$ & \red{0.98} 
        %& 88.70& 96.90 & 99.50 & 99.50 & 0.91
        \\ 
        & \textbf{SWAP (ours)} & \textbf{76.94}&	\textbf{71.21}	&\textbf{100.00} & \textbf{0} & \textbf{0.00} & 98.19&	\textbf{95.65}&	\textbf{100.00} & \textbf{0} & \textbf{0.00} & \textbf{96.16}&	97.43&	\textbf{100.00} & \textbf{0} & \textbf{0.00} 
        %& 90.60 & 91.35 & 99.97 & $10^{-3182}$ & 0.00
        \\ 

       \bottomrule
    \end{tabular}
    }
    
    \label{main_res_sota}
    %\vspace{-0.5em}
\end{table*}

\section{Additional Experiments}
\label{add_exp}

\subsection{Experiments on SOTA Prompt Tuning Method}
\label{sota_main}
We validate the robustness of our method by conducting additional experiments using the SOTA Prompt Tuning Method, ATPrompt \citep{li2024advancing}. Specifically, Table \ref{main_res_sota} presents the main performance results for SWAP in comparison with other baseline methods when using ATPrompt \citep{li2024advancing}. Furthermore, Table \ref{ind_res_sota} shows the verification outcomes of the watermarked model, independent prompts, and independent verification classes, all implemented using ATPrompt \citep{li2024advancing}. Collectively, these experiments strongly demonstrate that our proposed SWAP method maintains superior effectiveness and reliability even when integrated with the SOTA prompt tuning method.

\begin{table*}[t!]

    \centering
    \caption{The more main results with the watermarked model (Watermarked), independent prompt (Ind Prompt), and independent verification classes (Ind Veri) using ATPrompt as prompt tuning method.}
    % \vspace{-0.8em}
    % \resizebox{0.47\textwidth}{!}{
     \setlength{\tabcolsep}{3pt}
        \begin{tabular}{c|c|cc |cc |cc}
        \toprule
          & Dataset{$\rightarrow$}
         & \multicolumn{2}{c}{ImageNet}
         & \multicolumn{2}{c}{Caltech101} &
          \multicolumn{2}{c}{OxfordPets} 
         % &\multicolumn{5}{c}{Food101}
         \\
        
           \cmidrule(lr){3-4}  \cmidrule(lr){5-6} \cmidrule(lr){7-8} 
           %\cmidrule(lr){18-22}

        \makecell{Prompt Tuning \\Method{$\downarrow$}}& \makecell{Scenario$\downarrow$}
        &  WSR     &p-value  & WSR     &p-value & WSR     &p-value 
        % & Base    & Novel  & WSR     &p-value &$\hat{H}$
        \\
       \midrule
       \multirow{3}{*}{ATPrompt} 
        & Ind Prompt & 3.41& 1&2.15 &  1 & 7.69 & 1 
        %& 90.67& 91.53& 4.10&  0.70 & -
        \\ 
        & Ind Veri & 4.92 & 1 & 3.06 &  1 & 1.12 & 1 
        %& 0.70& 99.60 & 0.70&  0.70 & -
        \\ 
        & Watermarked & 100.00 & 0 & 100.00 &  0 & 100.00 & 0 
        %& 90.30& 91.10& 99.60 & 0.70& 0.91
        
        \\ 

       \bottomrule
    \end{tabular}
    % }
    \vspace{-0.8em}
    \label{ind_res_sota}
\end{table*}

\begin{table*}[t!]
    \centering

    \caption{The results of SWAP in comparison with independent prompt (Ind Prompt), independent verification classes (Ind Veri), and SWAP without verification classes (SWAP (no Veri)) across 11 datasets. HM represents harmonic mean.}
    % \vspace{-0.8em}
    \resizebox{\textwidth}{!}{
     \setlength{\tabcolsep}{3pt}
        \begin{tabular}{c|c|cccccc|ccccc|ccccc|ccc}
        \toprule
          & Scenario$\rightarrow$&
         \multicolumn{6}{c}{SWAP (ours)} & \multicolumn{5}{c}{Ind Prompt}& 
         \multicolumn{5}{c}{Ind Veri} &
         \multicolumn{3}{c}{SWAP (no Veri)}  \\
        
           \cmidrule(lr){3-8}  \cmidrule(lr){9-13} \cmidrule(lr){14-18} \cmidrule(lr){19-21}

        Dataset$\downarrow$& \makecell{Prompt Tuning \\Method$\downarrow$}& \makecell{ACC \\(Base)}   & \makecell{ACC \\(Novel)}& HM&WSR  &p-value&$\hat{H}$ & \makecell{ACC \\(Base)}  & \makecell{ACC \\(Novel)} &HM &WSR&p-value& \makecell{ACC \\(Base)}  & \makecell{ACC \\(Novel)} & HM& WSR&p-value& \makecell{ACC \\(Base)}  & \makecell{ACC \\(Novel)} & HM \\
       \midrule
        \multirow{3}{*}{\makecell{Average on \\11 Datasets}} & CoCoOp & 79.64 & 73.25 & 76.31 & 99.78 & 0 & -0.02 & 80.04 & 73.05 & 73.98 & 3.43 & 1 & 80.05 & 72.67 & 75.85 & 4.67 & 1 & 79.99 & 72.63 & 75.82 \\ 
& MaPLe & 82.52 & 73.81 & 77.92 & 99.91 & 0 & 0.01 & 82.74 & 74.90 & 78.38 & 2.77 & 1 & 82.33 & 72.98 & 77.02 & 4.09 & 1 & 82.67 & 73.60 & 77.55 \\ 
& PromptSRC & 83.46 & 75.50 & 79.28 & 99.91 & 0 & 0.01 & 83.80 & 75.86 & 79.41 & 2.84 & 1 & 83.62 & 75.30 & 78.99 & 4.37 & 1 & 83.66 & 76.01 & 79.42 \\ 
        \midrule
        \multirow{3}{*}{ImageNet} & CoCoOp & 75.89 & 70.10 & 72.88 & 99.92 & 0 & 0.00 & 76.10 & 70.58 & 73.24 & 4.27 & 1 & 75.52 & 70.85 & 73.11 & 1.30 & 1 & 75.95 & 70.92 & 73.35 \\ 
& MaPLe & 77.13 & 69.26 & 72.98 & 99.95 & 0 & 0.01 & 77.29 & 69.85 & 73.38 & 3.20 & 1 & 77.11 & 69.68 & 73.21 & 2.86 & 1 & 77.18 & 69.74 & 73.27 \\ 
& PromptSRC & 77.43 & 70.48 & 73.79 & 99.97 & 0 & 0.00 & 77.83 & 70.78 & 74.14 & 3.01 & 1 & 77.63 & 70.46 & 73.87 & 2.38 & 1 & 77.73 & 70.59 & 73.99 \\ 
\midrule
        \multirow{3}{*}{Caltech101} & CoCoOp & 97.65 & 93.10 & 95.32 & 99.34 & 0 & 0.01 & 97.73 & 94.57 & 96.12 & 2.62 & 1 & 96.64 & 94.76 & 95.69 & 1.20 & 1 & 97.56 & 94.81 & 96.17 \\ 
& MaPLe & 97.30 & 95.31 & 96.29 & 99.89 & 0 & -0.01 & 97.87 & 94.54 & 96.18 & 1.53 & 1 & 97.61 & 94.10 & 95.82 & 4.04 & 1 & 98.47 & 94.93 & 96.67 \\ 
& PromptSRC & 96.28 & 94.21 & 95.23 & 99.78 & 0 & 0.00 & 97.95 & 94.43 & 96.16 & 5.90 & 1 & 97.55 & 94.65 & 96.08 & 4.69 & 1 & 98.32 & 94.78 & 96.52 \\ 
\midrule
        \multirow{3}{*}{OxfordPets} & CoCoOp & 94.92 & 97.40 & 96.14 & 99.55 & 0 & 0.00 & 95.52 & 97.71 & 96.60 & 7.10 & 1 & 94.47 & 97.20 & 95.82 & 3.52 & 1 & 95.11 & 97.50 & 96.29 \\ 
& MaPLe & 95.03 & 96.82 & 95.92 & 99.89 & 0 & 0.01 & 95.22 & 97.99 & 96.59 & 3.30 & 1 & 95.48 & 97.48 & 96.47 & 3.69 & 1 & 95.64 & 97.83 & 96.72 \\ 
& PromptSRC & 96.13 & 96.80 & 95.96 & 99.94 & 0 & 0.01 & 95.67 & 97.53 & 96.59 & 1.40 & 1 & 95.64 & 97.82 & 96.72 & 6.82 & 1 & 95.49 & 97.87 & 96.67 \\ 
\midrule
        \multirow{3}{*}{\makecell{Stanford \\Cars}} & CoCoOp & 68.92 & 73.09 & 70.94 & 99.90 & 0 & 0.01 & 68.50 & 75.27 & 71.73 & 2.08 & 1 & 69.77 & 74.42 & 72.02 & 1.96 & 1 & 69.77 & 74.46 & 72.04 \\ 
& MaPLe & 75.25 & 73.09 & 74.15 & 99.98 & 0 & 0.01 & 75.89 & 73.78 & 74.82 & 2.28 & 1 & 73.96 & 73.19 & 73.57 & 1.96 & 1 & 75.91 & 73.18 & 74.52 \\ 
& PromptSRC & 76.37 & 75.03 & 75.69 & 99.88 & 0 & 0.00 & 77.83 & 74.80 & 76.28 & 4.33 & 1 & 76.66 & 75.17 & 75.91 & 3.76 & 1 & 76.73 & 75.42 & 76.07 \\ 
\midrule
        \multirow{3}{*}{Flowers102} & CoCoOp & 92.59 & 71.91 & 80.95 & 99.85 & 0 & 0.00 & 93.87 & 72.20 & 81.62 & 8.44 & 1 & 94.12 & 71.28 & 81.12 & 15.32 & 1 & 93.96 & 72.07 & 81.57 \\ 
& MaPLe & 96.74 & 73.05 & 83.24 & 99.93 & 0 & -0.01 & 97.09 & 72.38 & 82.93 & 0.92 & 1 & 96.49 & 72.98 & 83.10 & 0.14 & 1 & 96.85 & 73.51 & 83.58 \\ 
& PromptSRC & 97.75 & 76.10 & 85.58 & 99.93 & 0 & 0.00 & 97.43 & 76.74 & 85.86 & 0.07 & 1 & 97.82 & 76.95 & 86.14 & 0.85 & 1 & 98.08 & 75.74 & 85.47 \\ 
\midrule
        \multirow{3}{*}{Food101} & CoCoOp & 90.30 & 91.18 & 90.74 & 99.97 & 0 & 0.00 & 90.70 & 91.25 & 90.97 & 2.87 & 1 & 90.24 & 91.23 & 90.73 & 0.49 & 1 & 90.53 & 91.46 & 90.99 \\ 
& MaPLe & 90.32 & 91.33 & 90.82 & 99.90 & 0 & 0.01 & 90.18 & 91.61 & 90.89 & 0.43 & 1 & 90.25 & 91.25 & 90.75 & 3.30 & 1 & 90.62 & 91.67 & 91.14 \\ 
& PromptSRC & 90.60 & 91.35 & 90.97 & 99.93 & 0 & 0.00 & 90.80 & 91.67 & 91.23 & 4.21 & 1 & 90.86 & 91.80 & 91.33 & 7.29 & 1 & 90.99 & 91.85 & 91.42 \\ 
\midrule
        \multirow{3}{*}{\makecell{FGVC \\Aircraft}} & CoCoOp & 33.51 & 32.13 & 32.81 & 99.82 & 0 & -0.08 & 33.71 & 33.29 & 33.50 & 4.14 & 1 & 34.51 & 31.37 & 32.87 & 4.86 & 1 & 34.64 & 30.38 & 32.37 \\ 
& MaPLe & 37.60 & 31.61 & 34.35 & 99.58 & 0 & 0.04 & 38.54 & 34.73 & 36.54 & 6.90 & 1 & 38.30 & 31.49 & 34.56 & 3.06 & 1 & 36.81 & 31.69 & 34.06 \\ 
& PromptSRC & 40.54 & 37.19 & 38.79 & 99.88 & 0 & 0.01 & 41.40 & 37.67 & 39.45 & 0.12 & 1 & 41.42 & 37.19 & 39.19 & 2.82 & 1 & 40.23 & 36.86 & 38.47 \\ 
\midrule
        \multirow{3}{*}{SUN397} & CoCoOp & 79.56 & 76.69 & 78.10 & 99.94 & 0 & 0.00 & 78.40 & 78.21 & 78.30 & 1.16 & 1 & 79.36 & 76.80 & 78.06 & 8.51 & 1 & 79.39 & 76.92 & 78.14 \\ 
& MaPLe & 80.53 & 77.20 & 78.83 & 99.97 & 0 & 0.02 & 81.27 & 77.40 & 79.29 & 3.13 & 1 & 80.94 & 77.33 & 79.09 & 1.37 & 1 & 81.57 & 77.53 & 79.50 \\ 
& PromptSRC & 82.59 & 78.57 & 80.53 & 99.99 & 0 & 0.00 & 82.74 & 78.98 & 80.82 & 4.09 & 1 & 82.75 & 78.51 & 80.57 & 6.47 & 1 & 82.90 & 78.64 & 80.71 \\ 
\midrule
        \multirow{3}{*}{DTD} & CoCoOp & 74.65 & 59.18 & 66.02 & 99.63 & 0 & -0.03 & 77.18 & 56.04 & 64.93 & 3.02 & 1 & 76.42 & 57.25 & 65.46 & 2.05 & 1 & 77.48 & 57.21 & 65.82 \\ 
& MaPLe & 80.09 & 57.25 & 66.77 & 100.00 & 0 & 0.02 & 80.56 & 59.78 & 68.63 & 0.12 & 1 & 80.56 & 56.88 & 66.68 & 10.63 & 1 & 80.86 & 57.63 & 67.30 \\ 
& PromptSRC & 82.41 & 60.14 & 69.54 & 99.88 & 0 & 0.03 & 83.30 & 59.30 & 69.28 & 3.74 & 1 & 82.41 & 60.02 & 69.46 & 2.05 & 1 & 82.49 & 60.37 & 69.72 \\ 
\midrule
        \multirow{3}{*}{EuroSAT} & CoCoOp & 88.00 & 65.54 & 75.13 & 99.87 & 0 & -0.06 & 86.43 & 59.64 & 70.58 & 0.05 & 1 & 87.36 & 62.60 & 72.94 & 8.77 & 1 & 83.42 & 61.35 & 70.70 \\ 
& MaPLe & 93.90 & 71.23 & 81.01 & 99.94 & 0 & 0.02 & 91.98 & 73.41 & 81.65 & 0.41 & 1 & 91.40 & 63.44 & 74.90 & 8.65 & 1 & 91.56 & 64.92 & 75.97 \\ 
& PromptSRC & 92.69 & 72.39 & 81.29 & 99.97 & 0 & 0.02 & 90.80 & 74.59 & 81.90 & 2.64 & 1 & 90.29 & 68.74 & 78.05 & 7.77 & 1 & 90.48 & 72.73 & 80.64 \\ 
\midrule
\multirow{3}{*}{UCF101} & CoCoOp & 80.04 & 75.39 & 77.65 & 99.83 & 0 & -0.02 & 82.34 & 74.74 & 78.36 & 1.95 & 1 & 82.17 & 71.66 & 76.56 & 3.41 & 1 & 82.12 & 71.77 & 76.60 \\ 
& MaPLe & 83.87 & 75.77 & 79.61 & 99.95 & 0 & 0.03 & 84.23 & 78.47 & 81.25 & 8.22 & 1 & 83.51 & 75.01 & 79.03 & 5.30 & 1 & 83.94 & 77.01 & 80.33 \\ 
& PromptSRC & 86.29 & 78.24 & 82.07 & 99.89 & 0 & 0.01 & 86.00 & 77.93 & 81.77 & 1.78 & 1 & 86.81 & 77.01 & 81.62 & 3.19 & 1 & 86.79 & 81.29 & 83.95 \\ 

       \bottomrule
    \end{tabular}
    }

    \label{xalldataset}
\end{table*}

\subsection{Experiments on More Datasets}
\label{other_data}
We conducted extensive evaluations on 11 datasets, which covers a wide range of recognition tasks, including ImageNet \citep{deng2009imagenet} and Caltech101 \citep{fei2004learning} which consists of generic objects; OxfordPets \citep{parkhi2012cats}, StanfordCars \citep{krause20133d}, Flowers102 \citep{nilsback2008automated}, Food101 \citep{bossard2014food}, and FGVCAircraft \citep{maji2013fine} for fine-grained classification, SUN397 \citep{xiao2010sun} for scene recognition, UCF101 \citep{soomro2012dataset} for action recognition, DTD \citep{cimpoi2014describing} for texture classification, and EuroSAT \citep{helber2019eurosat} which consists of satellite images. All other settings remain consistent with our main experiments.

For evaluation, we use the harmonic mean (HM) between base and novel class accuracy, which represents generalization performance. In addition to SWAP, independent prompt, and independent verification classes scenarios, we tested the accuracy of SWAP under normal usage conditions, \ie, SWAP's accuracy when verification classes are not included in the test classes, denoted as SWAP (no Veri).

As shown in Table \ref{xalldataset}, Experimental results demonstrate that our method maintains high accuracy on both base and novel classes across all datasets, while simultaneously achieving high WSR and small p-values. In the SWAP (no Veri) scenario, both base and novel class accuracies remain high, indicating that our method does not impair normal user experience. In independent prompt and independent verification classes scenarios, the WSRs are all below 10\% and the p-values are relatively large across all datasets.

\section{Reproducibility Statement}
The detailed configurations of datasets, models, hyperparameters, and computational resources are provided in Section~\ref{main_settings} and Appendix~\ref{set_main}. The full implementation of our methods (including codes and model checkpoints) will be released upon the acceptance of this paper.




\end{appendices}


% \bibliography{sn-bibliography}% common bib file
% %% if required, the content of .bbl file can be included here once bbl is generated
% %%\input sn-article.bbl

\end{document}